# The effects of turbulence generated in Big Bang nucleosynthesis

Charles Francis

*C.E.H.Francis.75@cantab.net*

ABSTRACT

The continuity equation requires that energy released in nuclear fusion flows away from the point of interaction and is not immediately thermalised into the CMB. It is seen that consequently the bulk of thermo-nuclear energy released in BBN is converted by radiation pressure to kinetic energy of gas motions and generates instability in homogeneous initial conditions. High levels of turbulence follow, triggering the Jeans mechanism and leading to the formation of galaxies and large-scale structure in the order of 10 Myrs. Friedmann cosmology is perturbed, creating a fractal Voronoi matter distribution and providing energy for reionisation. In keeping with observations, general relativity predicts that there is no bound on the size of the structures produced. The true (overall) rate of expansion is substantially below the observed (local) value of Hubble's constant, and is consistent with published estimates from the CMB for void models. The CMB is predicted to show near flatness over the observed region. It is seen from observation of fluid flows, from established physics, and from numerical solution of the Navier-Stokes equations that bisymmetric spiral vortices form where gas flows meet. These vortices necessarily have flat rotation curves at large radii and create conditions for spiral galaxy formation. It is shown by N-body simulation that enduring Sa, Sb and Sc spiral galaxies result from appropriate initial conditions. The simulations and analysis of spiral potential show that orbital velocities are greater than would be the case in an axisymmetric potential.

## 1 Introduction

In recent years there has been considerable interest in whether the Earth may reside in an underdense region of the universe, or void, as an alternative to Λ-CDM cosmology. Modelling has shown that such a possibility cannot be excluded on grounds of the fit to the CMB power spectrum, to baryon acoustic oscillation data or to supernova data as void models can produce at least as good a fit (e.g. Clarkson & Regis 2011, Nadathur & Sarkar 2011, Biswas, Notari & Valkenburg 2010), but a realistic mechanism breaking the symmetry described by the cosmological principle while preserving the homogeneity and isotropy of initial conditions has been lacking. Likewise, standard cosmology lacks a mechanism to explain observed voids (e.g. Hunt & Sarkar 2010).

The standard arguments for the cosmological principle justify the solution of the Friedmann equation as a valid approximation on sufficiently large distance scales, but it will be seen here that local effects cannot be ignored. The equilibrium described by initial conditions of homogeneity and isotropy is necessarily unstable and leads to turbulence in baryonic gas. This paper will show that turbulent dynamics in a baryonic universe perturbs Friedmann cosmology to create structures consistent with observation at all distance scales.

It has been a problem for a baryonic model that variations in initial conditions appear too small to instigate gravitational collapse within a sensible timescale. This overlooks motions generated by the release of energy in matter-antimatter annihilation and Big Bang nucleosynthesis (BBN), which dominated over *localised* gravitational effects by many orders of magnitude. It has been suggested that the energy of fusion released in nucleosynthesis can be ignored compared to the high background energy in the radiation dominated phase of the early universe. This is true of overall behaviour (solution of the Friedmann equation), but has no bearing on the dynamics considered in this paper. Local dynamics depends on local energy *differences* and is not altered by high background energy.

It is also suggested that energy from fusion was rapidly thermalised into the microwave background. This violates the continuity equation for mass-energy; energy released locally cannot immediately be redistributed uniformly in the background, but must flow from the point at which it is released. As seen in section 2, a small overdensity in mass increases the rate of fusion. Energy released must be radiated away. Since the flow of energy is momentum, radiation pressure generates gas flows and necessarily leads to instability caused by escalating differences in mass densities. Thus it will be seen that the release of nuclear energy during BBN was such as to generate massive high velocity flows. Colliding residual flows following BBN then generated fronts of high pressure and density from which gravitational collapse could proceed within a very short cosmological timescale, in accordance with the Jeans mechanism.

In contrast to special relativity, in which speeds are bounded by the speed of light, in general relativity gravitational collapse means the contraction of space itself. Expansion takes place more rapidly in less dense regions, opening up voids, while contraction takes place in denser regions under gravitational collapse. There is no bound on the rate of local contraction of space during gravitational collapse.

According to Newton's shell theorem, mass outside of a collapsing region has little effect on the collapse (it would have no effect under conditions of spherical symmetry; although this argument is suspect in a Newtonian universe where it is described as "the Jean swindle" because the divergence of integrals over all space leads to inconsistencies, it is justified in an expanding universe because gravitational effects propagate at the speed of light, and only gravitating mass within the light cone need be considered, Falco et al. 2013). Then, collapse under the Jeans mechanism takes place identically on all distance scales in approximately equal time. Since there is no bound on the rate of expansion and contraction of space, the only bound on the sizes of voids and structures created in gravitational collapse is the size of the universe itself. As described below, structures observed at all distance scales, from planar alignments of satellites of the Milky Way to supervoids and great walls (including large quasar and gamma ray burster groups if confirmed) may be remnants of intense activity in the very early universe.

The principle historical evidence for cold dark matter has been derived from gravitational rotation curves. This will be reconsidered here using N-body simulations taking account of spiral potential and gas in the intergalactic medium. It is found that cold dark matter is not required to model spiral galaxies.

The virial theorem cannot be used to assert the existence of dark matter in clusters because the prediction in a Voronoi model is that clusters are not in dynamical equilibrium. Indeed, there is no good evidence to suggest that they should be gravitationally bound; in a baryonic universe they are not gravitationally bound.

Other evidence has come from the CMB, but void models have shown that other fits are possible in models with a lower overall rate of expansion.

Weak lensing measurements, for example on the Bullet cluster, have also shown the presence of dark matter (Clowe, Gonzalez & Markevitch 2004, Clowe et al. 2006), but Lee and Komatsu (2010) found the evidence incompatible with Λ-CDM cosmology. It is possible, for example, that dark matter in galaxy clusters revealed by weak lensing is accounted largely by neutrinos.

## 2 Primordial conditions

At $t = 100\,\mathrm{s}$, $z \approx 5 \times 10^8$, $k_\mathrm{B}T = 0.115$ MeV, radiation energy density was more than one million times greater than baryon density. The temperature of baryonic matter was thus stabilised to the radiation background, by Coulomb scattering between nucleons and electrons and by Compton scattering between electrons and photons. Consequently, apart from cooling due to the expansion of the universe, processes were isothermal.

The helium abundance shows that the release of thermonuclear energy due to BBN was greater by one to three orders of magnitude than the total released in all nuclear processes during 14 Gyrs since. If, due to an initial random fluctuation, a region was slightly overdense then the rate of interaction in that region was faster. While processes were isothermal, energy released in fusion cannot be directly thermalised into the microwave background because it is localised; the continuity equation for mass-energy ensures that it must flow away from the region where it is released. The flow of energy is momentum. Thus, the energy, 17.59 MeV, from each fusion process was transmitted away from the dense region as radiation, either in the form of direct products of fusion or from Compton scattering from energetic charged particles. The mean free path for a photon was in the order of kilometres. This energy was absorbed by surrounding gas, transferring momentum to the gas and generating a radial flow of gas away from the dense region.

Since the bulk of radial radiation was absorbed, the conversion of fusion energy to kinetic energy in bulk motions of gas was almost 100% efficient. The radial flow of gas away from the initially dense region then led to underdensity of that region. Another region would then become overdense where gas flows collide, creating an explosive front. Thus, even in the radiation dominated era, burning necessarily generates instability (just as for all observed uncontrolled burning processes). Each explosive front will have involved more matter and generated greater heat and greater flows, leading to escalating explosive forces in the each iteration of an essentially chaotic cycle. If, as the universe expanded, interactions took place in fronts, with relatively cool regions between the fronts, then, following an initial period of homogeneity, only a proportion of matter would have been interacting at any one time. BBN will have continued within fronts for longer than usually estimated, suggesting also a lower rate of expansion.

The total thermonuclear energy released in BBN was equivalent to kinetic energy of motion at 6% of the speed of light. The mechanism described above converts the major part of this to kinetic energy. Prior to BBN, even greater energy was released in matter-antimatter annihilation. As for fusion in BBN, matter-antimatter annihilation was faster in denser regions, creating explosive forces and flows away from denser regions leading to new explosive fronts where flows meet. We can thus be certain that the dynamics of the early universe was characterised by flows moving at a significant proportion of the speed of light (in this context, even 0.1% can be considered significant, representing less than 0.03% of the energy released, although this is certainly an underestimate). The high speeds generated will have ensured substantial mixing. Together with the isothermal conditions in a radiation dominated universe, the overall picture of homogeneity seen at decoupling in the cosmic background radiation is preserved. Although the distribution may appear spacially homogeneous, it does not follow that dynamical structure was uniform, at least on the scale on which it can be observed.

It has been argued that diffusion damping (Silk 1968) removes small scale (galaxy sized) static inhomogeneities in the radiation dominated era, but it cannot alter the bulk flow of momentum. Collisions between fronts cause high densities, which trigger fusion and fuel an ongoing process, terminating only when fronts contain sufficient matter to be gravitationally bound. At the energies involved, the escalation of these flows would have proceeded extremely rapidly, and would have generated the beginnings of large-scale structure.

As the universe expanded, the distances travelled by each flow increased, as did the size of each front where flows collide. An order of magnitude estimate for the eventual size of fronts can be given by considering an effect propagating at 1% of the speed of light. This would have a diameter in the order of 10 Mpc in comoving coordinates at the time of the formation of the cosmic microwave background, about five orders of magnitude less than can be detected with the Planck telescope.

Once inhomogeneities had arisen, gravitational collapse became the dominant mechanism for structure formation, as in standard models. Collapse is seeded by inhomogeneities in density and velocity generated by turbulent processes from BBN. Under the action of gravity, matter falls away from less dense regions, opening up voids, while regions of greater density collapse under the Jeans mechanism and are attracted towards each other. As voids open up, random variations seed collapse on greater distance scales, generating much larger structures than described by the initial granularity, and in accordance with observation.

As gases cooled due to the expansion of the universe, collisions and the effect of expansion itself damped out extreme flows quickly (on cosmological timescales), but residual motions continued for some time. Collisions between gas flows continued to create regions of high density, seeding collapse by the Jeans mechanism. If this was the first stage of galaxy formation and the emergence of large-scale structure, the earliest structures and population III stars will have formed in a time period in the order of 10 Myrs after the Big Bang (a typical free fall time for the Jeans mechanism), at $z$ in the order of 100. In practice, we see the first evidence of structure formation in the period of reionisation, starting about 100 Myrs later, at $z = \sim 20$.

In gravitational collapse to a cellular structure, rates of change of proper distances are governed by geometry, not by the speed of light. The speed of light is a bound on local velocity, *not* on the rate of expansion of the universe following the Big Bang, or on the rate of change from a homogeneous beginning to a cellular geometry. In the current universe, we see expansion between galaxies at cosmological distances, but no expansion in gravitationally bound regions such as galaxies. Similarly, in a cellular matter distribution, expansion is faster in voids, and contraction takes place in regions collapsing to form walls. Changes in geometry during structure formation are thus highly inhomogeneous and anisotropic. Owing to the scale invariance of Jeans collapse, there is nothing to prevent the formation of voids and walls on scales of Gpc and greater within a short period of cosmological time following the Big Bang (tens of Myrs). It is thus reasonable to think that we may be in a void and that the largest structures, and the greatest mass, may lie beyond the horizon.

When structures formed along fronts between colliding gas flows were dense enough to be gravitationally bound, they will not have cooled with cosmological expansion. Gases in the intervening space will have cooled. Cooled atomic hydrogen and helium will have been gravitationally drawn to primordial structure, where it will have reionised on meeting the wall, in consequence of the loss of kinetic energy gained from falling from regions of lesser to greater density. Thus the energy for reionisation was essentially gravitational, as it is today in the warm-hot intergalactic medium which remains ionised due to the energy of infalling gas. Reionisation at $\sim 20 > z > \sim 6$, ~150 to ~750 Myrs after the Big Bang marks the major period of large scale structure formation as gas fell from voids into walls.

It is generally true that the mean results in turbulent systems are close to results predicted by calculations based on uniformity. Consequently little change is expected to the calculation of abundances of light elements. Detail effects of extreme fluctuations in density and rate of interaction during BBN are not obvious and it would be interesting to study whether the lithium deficit is accountable by variations in density and rate of interaction due to turbulence during BBN. With larger telescopes in the future it may be possible to determine whether there is a difference between abundances in great walls and those in more rarefied regions.

## 3 Fractal Voronoi tessellation

Zel'dovich (1970) observed that in a universe in which structure formation is dominated by baryonic matter gas pressure resists collapse towards a point. He predicted that gravitational collapse of gas tends to generate flattened walls. Further studies have shown that the resultant structure forms a network of cells, or voids, separated by flat walls (Einasto, Joeveer and Saar 1980, Gurbatov, Saichev and Shandarin 1989), and that matter ultimately congregates in filaments, where the walls meet (Hoffman, Salpeter and Wasserman 1983, Bertschinger 1985). The resulting configuration can be likened to a Voronoi tessellation defined by the perpendicular bisectors of "seeds" (figure 1, and Peacock 1998 figure 16.9). Gaseous matter falls away from regions of low density represented by the seeds. This mechanism is retained in the model considered here, but the walls, or dense regions, are initially created by colliding gas flows, which then seed further gravitational collapse. In keeping with observation, large voids or cells in the galactic foam are expected typically to have an approximately spherical form.

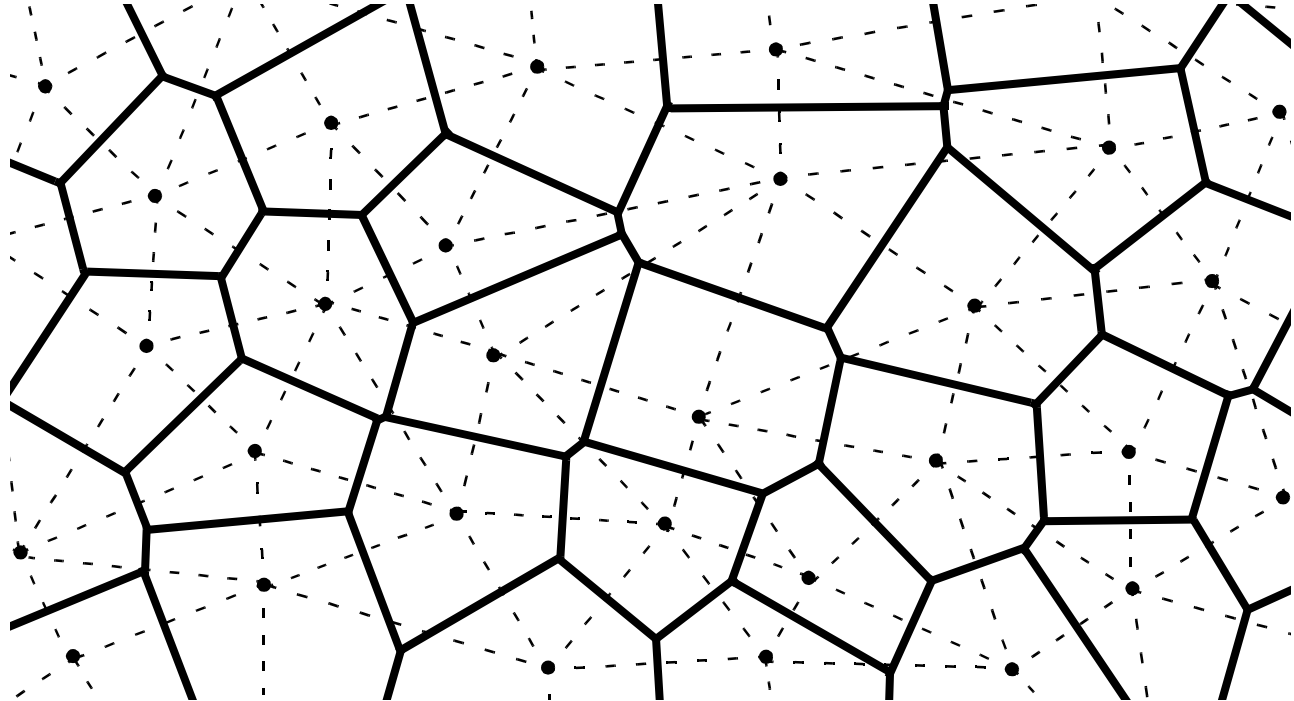

**Figure 1:** In Voronoi tessellation, perpendicular bisectors are drawn between "seed" points, to divide a domain into regions in which each point is closer to the seed for that region than any other seed. This picture is justified because, in a first approximation, Newton's shell theorem means that the gravity of surrounding matter has little overall effect. Gravity is thus better understood as a net force away from regions of low density than as a force towards high density.

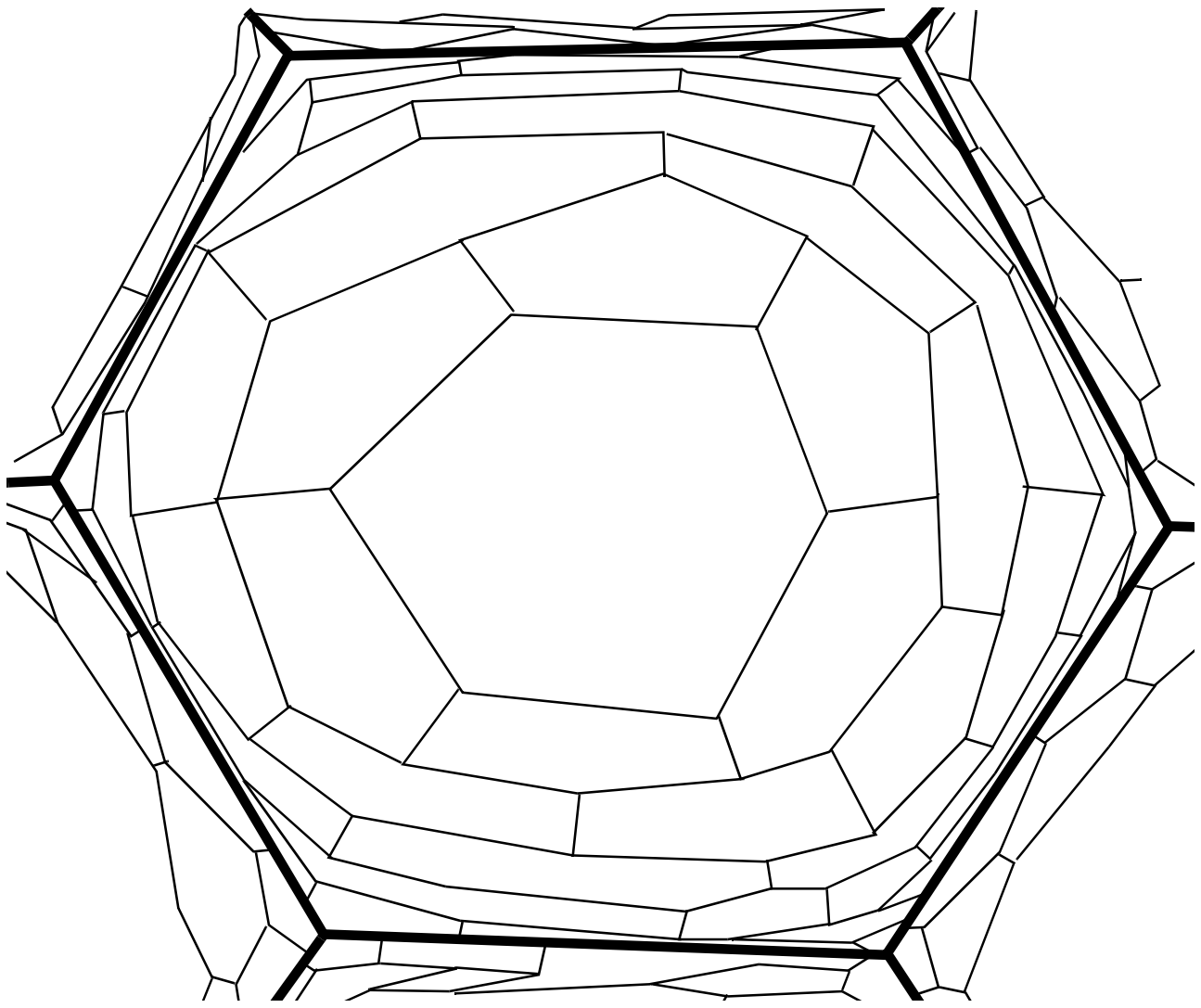

**Figure 2:** Fractal Voronoi tessellation. Seeded by dense regions resulting from colliding gas flows, gas from the Big Bang falls into walls. Since time for collapse depends on density but not on distance, it breaks up into smaller walls, just as gas clouds break up into stars. Under baryonic collapse a cellular structure is expected to be repeated over a wide range of distance scales.

It is well established that the free fall time, $(G\rho)^{-1/2}$, of the Jeans mechanism depends on density, but is independent of the size of the region of collapsing gas. This is the reason collapsing gas clouds break up to form many new stars. Turbulence will also have had a role in breaking up gas clouds, by causing rapid variations in pressure and flow. Random variations in the size of voids seed collapse on greater distance scales. Consequently, collapse takes place on all distance scales, creating large cells separated by large walls and simultaneously dividing the large cells into smaller cells separated by smaller walls. We should thus expect a fractal structure to emerge; each large cell is divided into smaller regions, which are themselves further divided into smaller cells (figure 2). The larger walls themselves contain smaller walls, generating a layered structure within the walls.

In practice, we observe alignments on all scales from great walls at distances of hundreds of megaparsecs, to the galactic foam first identified by de Lapparent, Geller and Huchra (1986, figure 3) and confirmed in subsequent surveys, to the Local Sheet containing the Council of Giants (McCall 2014, figure 4) down to the alignments of old globular clusters in planes through the Milky Way (Lynden-Bell 1975, Pawlowski, Pflamm-Altenburg and Kroupa 2012), Andromeda (Ibata, Lewis, Conn et al. 2013, Shaya and Tully 2013), and Centaurus A (Tully et al., 2015). The satellites in these structures are orbiting in the same direction, as one expects if they are formed from gas motions in walls of the foam. This does not mean that the structure should be in a stable orbit about the galaxy, but reflects the motions of gases within the wall from which the satellites were formed.

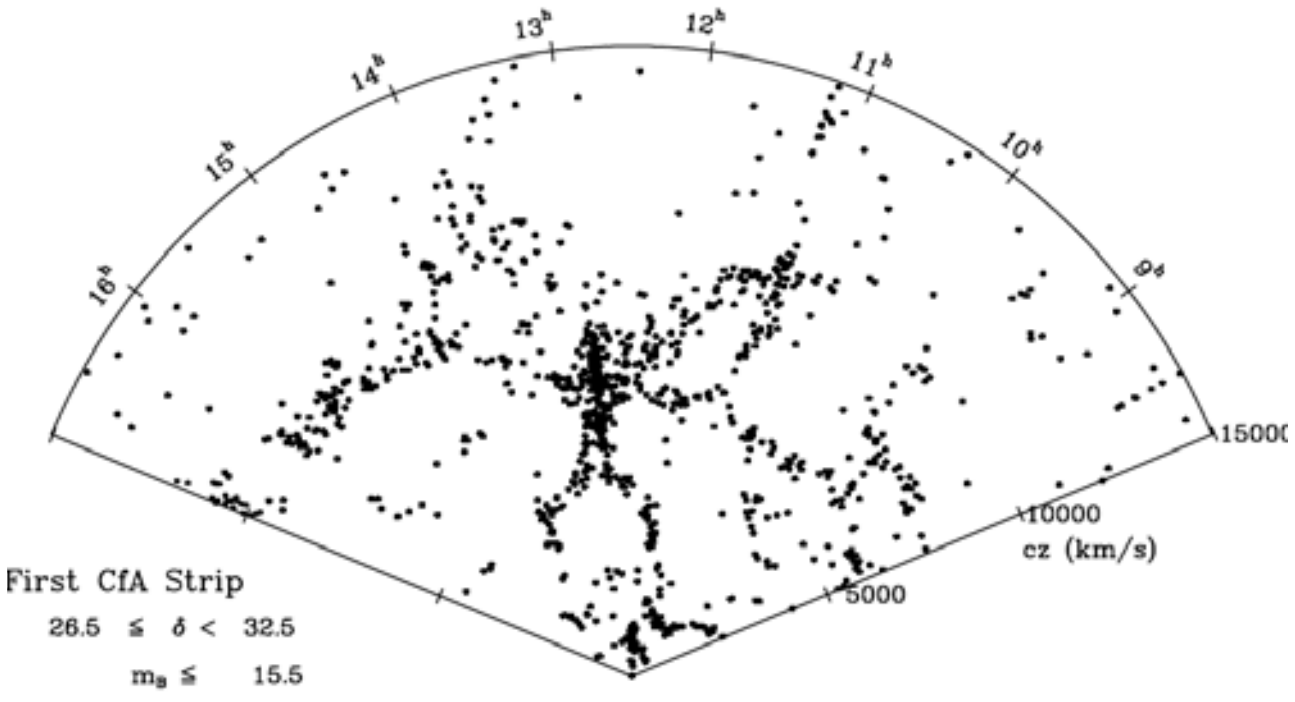

**Figure 3:** Clustering of galaxies in the first CfA strip, from Geller, Huchra & de Lapparent, Proceedings of the AU Symposium, Beijing, People's Republic of China, 1986. Image copyright, the Smithsonian Astrophysical Observatory.

On still larger distance scales, claimed observations of large quasar groups, the giant GRB ring (Balázs et al. 2015) and the Great GRB wall (Horvath, Hakkila & Bagoly 2014) suggest that a pattern of inhomogeneity may continue at least as far as redshift 2. It is to be hoped that the next generations of larger telescopes will enable us to confirm whether these structures are indeed indicators of inhomogeneity at greater distance scales.

As dense fronts between gas flows break up under the Jeans mechanism into galaxies, galaxy groups with similar motions were created on the fronts between different flows. Such comoving groups are observed in practice and are not explained in standard cosmology. They do not arise naturally in hierarchical models.

Clusters are formed at the intersections of walls. Galaxy clusters contain galaxies created by different flows, and which have been drawn from the walls to the intersection. In keeping with observation, high velocity dispersions are predicted, since there is no significant damping mechanism on the motions of galaxies and no mechanism according to which galaxies in clusters would be gravitationally bound. Clusters are observed at meeting points between walls because intersections between walls are expected to be regions of high density, because material is drawn to this region from the wall and from the void; galaxies continue to join the cluster from the wall as others leave. In fact, given scale of intergalactic distances, the timescale for the dispersion of galaxies in a typical cluster of diameter 2 to 10 Mpc is in the order of the age of the universe or greater.

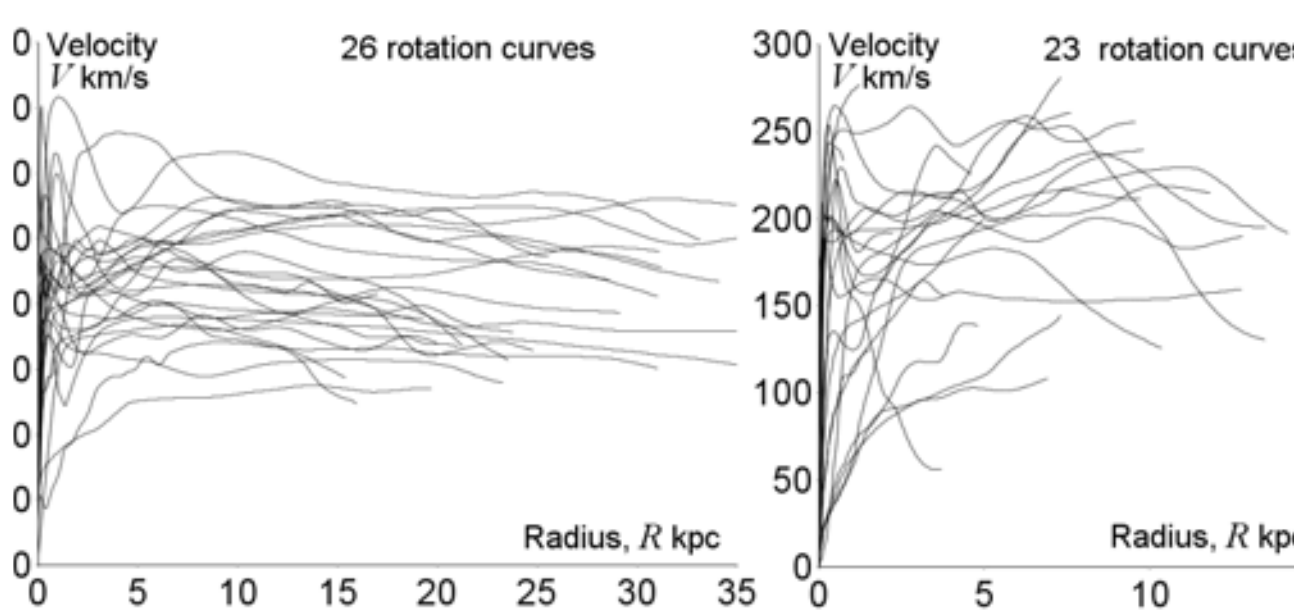

**Figure 5:** Rotation curves of spiral galaxies obtained by combining CO data for the central regions, optical for discs, and H I for outer disc and halo. Larger galaxies are shown on the left (data from Sofue et al., 1999).

## 4 Rotation curves

As observed in the Hubble Deep Field, early galaxies were small. Simulations show that gas accretion was the primary process of galactic growth even in models seeded by dark matter (e.g. Keres et al. 2005, Benson & Bower 2011). Direct evidence for gas accretion has been observed (Lehner et al. 2013, Crighton et al. 2013). Because gas merges from all sides in a supercluster there is little net overall rotation; we observe mainly elliptical galaxies in supercluster cores. Smaller amounts of gas left behind in the collapse are more likely to acquire angular momentum. Rotating gas clouds in outer regions of clusters form into spiral galaxies, such as the Milky Way. Rotation was an important mechanism in the development of thin discs and spiral structure after the original process of galaxy formation, but, as shown by alignments in the galactic foam, galactic planes may have been established from the collapse of gas prior to the formation of galaxies themselves.

As described in a review by Sofue and Rubin (2001), ignoring peculiarities such as lopsidedness and small numbers of rotation curves for which velocities decrease significantly at high galactic radius, typical rotation curves for spiral galaxies have common characteristics. Atypical features are not correlated with any other property, such as morphology, luminosity, or local galaxy density. The rotation curves for most luminous spirals follow a similar pattern, rising

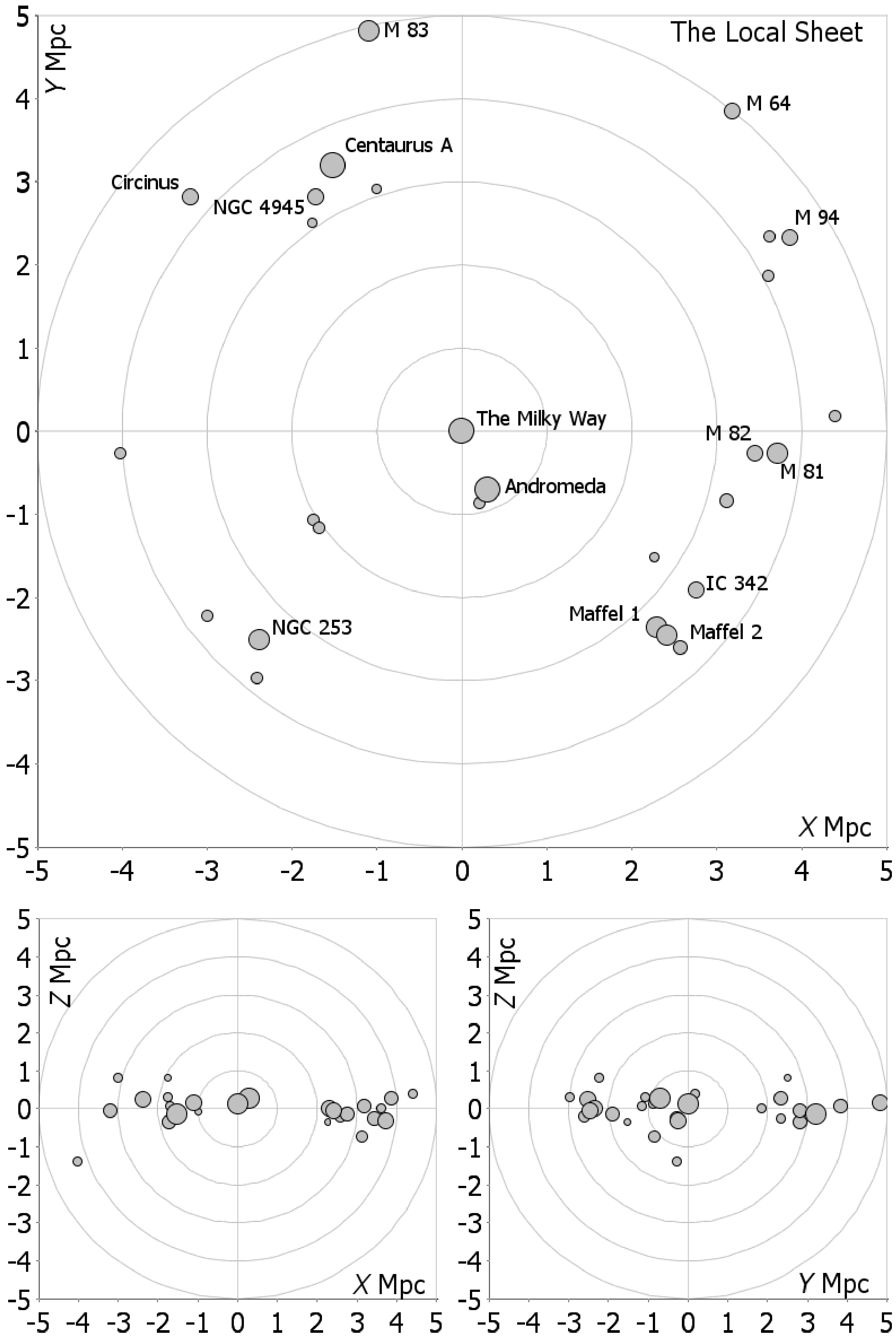

**Figure 4:** Top and side views of the local sheet, showing the 28 brightest galaxies within 5Mpc, and the 14 members of the Council of Giants. Data from McCall M.L., 2014, MNRAS, 440, 405-426

to a steep peak close to the galactic centre, followed by trough, and rising again to a broad maximum in the disc before falling off gently in the outer region (figure 5). The curves of small galaxies and low-surface brightness galaxies tend to rise throughout the range of measurements. There are exceptions. In some galaxies the rate of rotation declines rapidly in the outer region, sometimes declining even faster than predicted by Newtonian gravity. The existence of exceptions may be at least as important as the general pattern of rising curves, since they directly refute any explanation in terms of a universal gravitational law. In particular, the existence of curves declining faster than the Newtonian prediction directly refutes explanations both in terms of cold dark matter and in terms of modified gravity.

Standard interpretations of galaxy rotation curves have assumed a) that a galaxy can be treated in isolation, b) that gas in the vicinity of a galaxy is in a state of dynamical equilibrium with gravity, c) that gas is in near circular motion at any distance from the galactic centre, and d) that the density of gas declines with distance from the galactic centre. These assumptions are not necessarily valid. Gas is depleted by star formation within galaxies. Circular motion is an extremum of minimum energy, and is unlikely for systems on the scale of galaxies and greater where there are neither the timescales nor the damping processes to allow dynamical equilibrium to arise. The existence of galaxy groups with similar motions refutes the notion that galaxies can be treated in isolation.

In outer regions, gas is continuously accreted from the intergalactic medium, which forms walls between voids. Away from dense regions, where the intergalactic medium is a warm-hot plasma, the intergalactic medium is cool and difficult or impossible to detect. As it falls towards a galaxy, and meets resistance from the gas already in the galaxy, it begins to warm and we detect H I lines. Gas is unlikely to fall radially into a galaxy from the wall. Motions will have been imparted to gas within the wall by large-scale effects, starting from conditions following BBN and perturbed by gravitational effects and pressure. A rotational component is expected. As rotating gas falls into the gravitational well generated by a galaxy, its transverse orbital velocity will naturally increase, in consequence of conservation of angular momentum, and in agreement with the general form of typical curves. We may therefore expect to find gas in outer regions of galaxies rotating typically faster than the speed of circular motion under gravity alone, as is observed.

The rotation of a galaxy will have been determined by the rotation of gas from which it is formed, and will be correlated with current motion of surrounding gas. A substantial proportion of rotation curves, around half, show abnormalities, such as counter-rotation, lopsidedness, dips in rotation velocity, and rapidly falling curves in the outer region. Abnormalities cannot be explained using either modified gravity or cold dark matter, but they can be expected to occur frequently if rotation curves are affected by the behaviour of infalling gas from the intergalactic medium, since they arise from changes in the velocities of surrounding gas since the time of formation.

The Milky Way rotation curve appears to fall off a little in outer regions and then continues roughly flat to a distance of 1Mpc (Sofue 2012), showing directly that the curve is not due only to the gravity of the Galaxy, but is also dependent on gas motions across a wider structure.

Thus, the outer part of a rotation curve, seen in gas but outside of the diameter of visible stars, does not describe the kinematics of the galaxy itself, but shows the motion of gas outside the galaxy. This gas is not in dynamical equilibrium with the gravity of the galaxy, but is subject to pressure from intergalactic gas. We should conclude that galaxy rotation curves do not offer evidence for exotic physics, but instead afford much more mundane (but, nonetheless, interesting) insights into the dynamics of intergalactic gas and the processes of galaxy evolution.

## 5 Formation of spiral galaxies

The fact that galaxies in the Local Sheet have similar peculiar motions and are aligned to a plane makes clear that they are formed from a single large scale process. Such structures can form when large scale gas flows meet along dense fronts, and can be expected consequent on the collapse of baryonic matter under gravity. Baryonic collapse generates a cellular structure, with matter tending to congregate at intersections between the walls, which are the primary regions for galaxy formation. At least in certain cases, we may treat the problem of galaxy formation as two dimensional. Baryonic matter collapses first to a two dimensional surface, comprising the wall of a cell. Gaseous matter in the wall then collides with gas with a different motion in the wall, possibly at an intersection with a smaller wall (this is indicated by this the alignments of old globular clusters in a plane through the Milky Way).

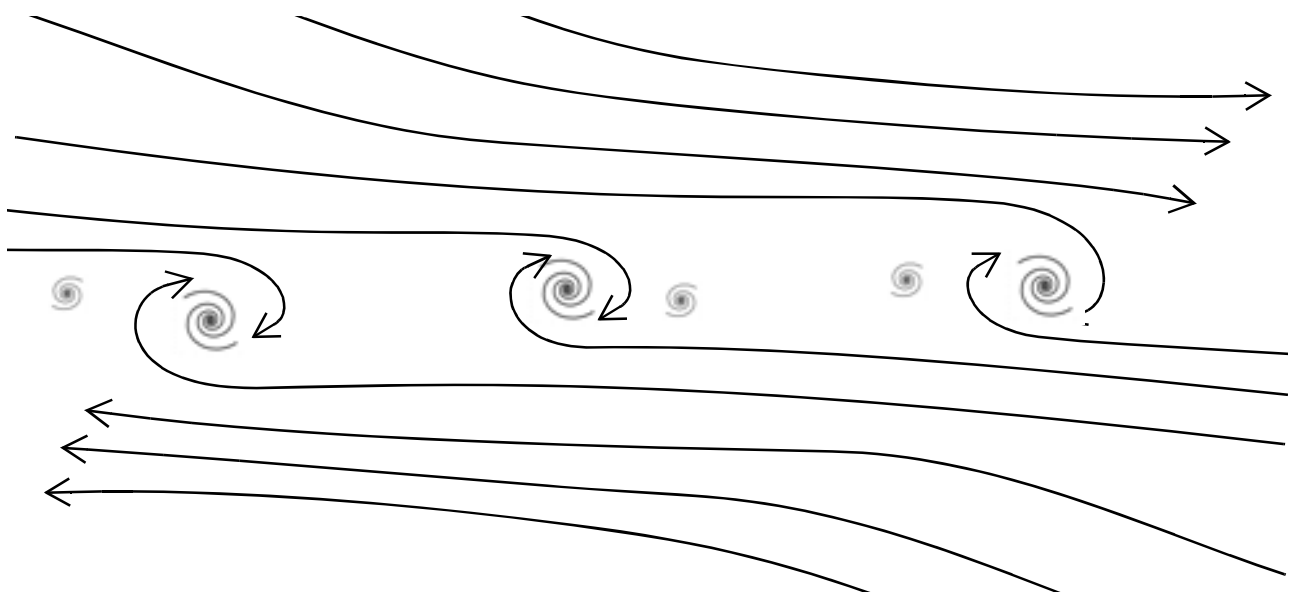

**Figure 6:** In oblique collision between gas flows, the flows are redirected and turbulence results along a front between the flows, creating conditions for formation of groups of spiral galaxies aligned in planes between the flows. Smaller vortices created by turbulence will generate smaller spiral galaxies (figure simplified for clarity).

When large-scale gas flows collide at an oblique angle, they will be redirected along the direction of the front between the flows. Turbulence will be generated between the flows, creating vortices from which spiral galaxies may form (figure 6). A typical spiral galaxy will then be found sandwiched between flows. In the outer region of the galaxy, and beyond the galactic diameter, the rate of rotation of the vortex is determined by the difference of the two gas flows. Planar flows will determine a flat outer curve. Given the distance and time scales for cosmological processes, the boundary condition is likely be similar to that at time of creation, but we may also expect a significant numbers of changes to the flow evidenced by the high proportion of curves with abnormalities.

Turbulence is characterized by chaotic properties, such as rapid variations of pressure and flow and the appearance of vortices on many distance scales often interacting with each other. The same will be true the intergalactic medium, where low densities lead to turbulence on large distance scales. In the vicinity of a large vortex there will also be a chaotic arrangement of smaller vortices leading to the formation of satellites. M51 is a classic example of a galaxy created from interacting vortices.

The Navier-Stokes equations, together with the continuity equations and (for compressible media) the gas laws, govern all fluid flows, including turbulent flows. In practice, exact digital modelling using the Navier-Stokes equation is possible only for laminar flows, due to the extreme changes in density and velocity generated by turbulence. Nonetheless each part of a turbulent flow can be considered as a laminar flow on a small enough distance scale and in a small enough time interval.

## 6 Modelling the formation of a spiral vortex

I modelled the evolution of a vortex surrounding a gravitating body sandwiched between colliding flows by initialising a rotating flow at uniform pressure. This is not intended as a physical initial condition, but the evolution shows where regions of high pressure and local turbulence arise under physical conditions, and indicates where star-formation is predicted to begin.

Because of the difficulty of detecting the cold neutral medium in intergalactic space it is not possible to determine exact parameters from observation. In practice a wide range of parameters are expected in nature and produce similar results. For definiteness, I modelled an isothermal process for a cold ideal gas with parameters approximate to the early universe, containing $5{\times}10^3$ particles per $m^3$ at a temperature 300K and consisting of six parts molecular hydrogen to one part helium. I modelled flows colliding at an angle 40° to the line of collision (figure 7 and figure 8). Changing these parameters altered the timescale for evolution, usually shortening it, but made no significant qualitative difference to the plots. I used a two dimensional model of the flows in a gaseous wall between cells, since the profile of the gas flows is unknown and seems unlikely to make a qualitative difference to results. I damped out velocity differences greater than $10^7$ ms$^{-1}$ between adjacent grid points, indicating the presence of extreme turbulence. This modestly increased the period of the simulation before it broke down due to high levels of turbulence and compounded digitisation errors (more sophisticated treatments are possible but this was adequate for purpose).

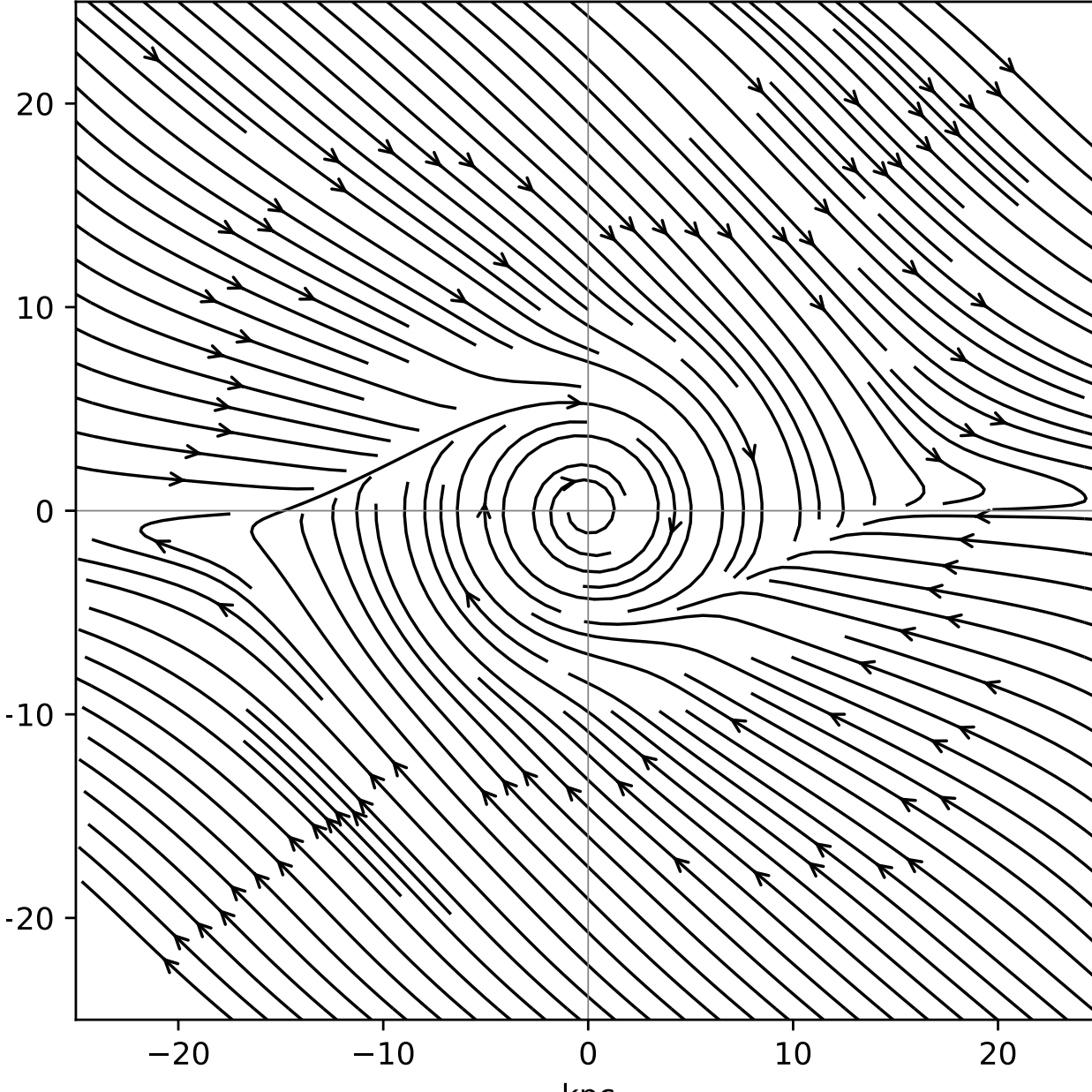

**Figure 7:** Streamline plot for gas 35 Myrs from initialisation found from digital solution of the Navier-Stokes equation. Gas streams collide at an angle 40° to the line of collision and a vortex forms around a gravitating mass. Changing parameters makes little difference to qualitative results.

The simulation ran for a period of 35 Myrs from initialisation before breaking down due to turbulence generated in high pressure regions. The streamlines for gas flow at 35 Myrs (figure 7) are not greatly changed from the initial condition. At this point peak pressure/density had increased by a factor of about 20 from the initial condition, indicating that star formation will start in dense regions either side of the vortex (figure 8 top). The wispy form of the high pressure region is notable since such forms are often seen in the vicinity of galaxies, where they are sometimes described as “tidal tails”. Tidal forces are extremely small, being a perturbation of the gravitational force, which is already small compared to the electromagnetic force. It may therefore be thought that high pressures generated by collision between gas flows is a more realistic mechanism for these phenomena.

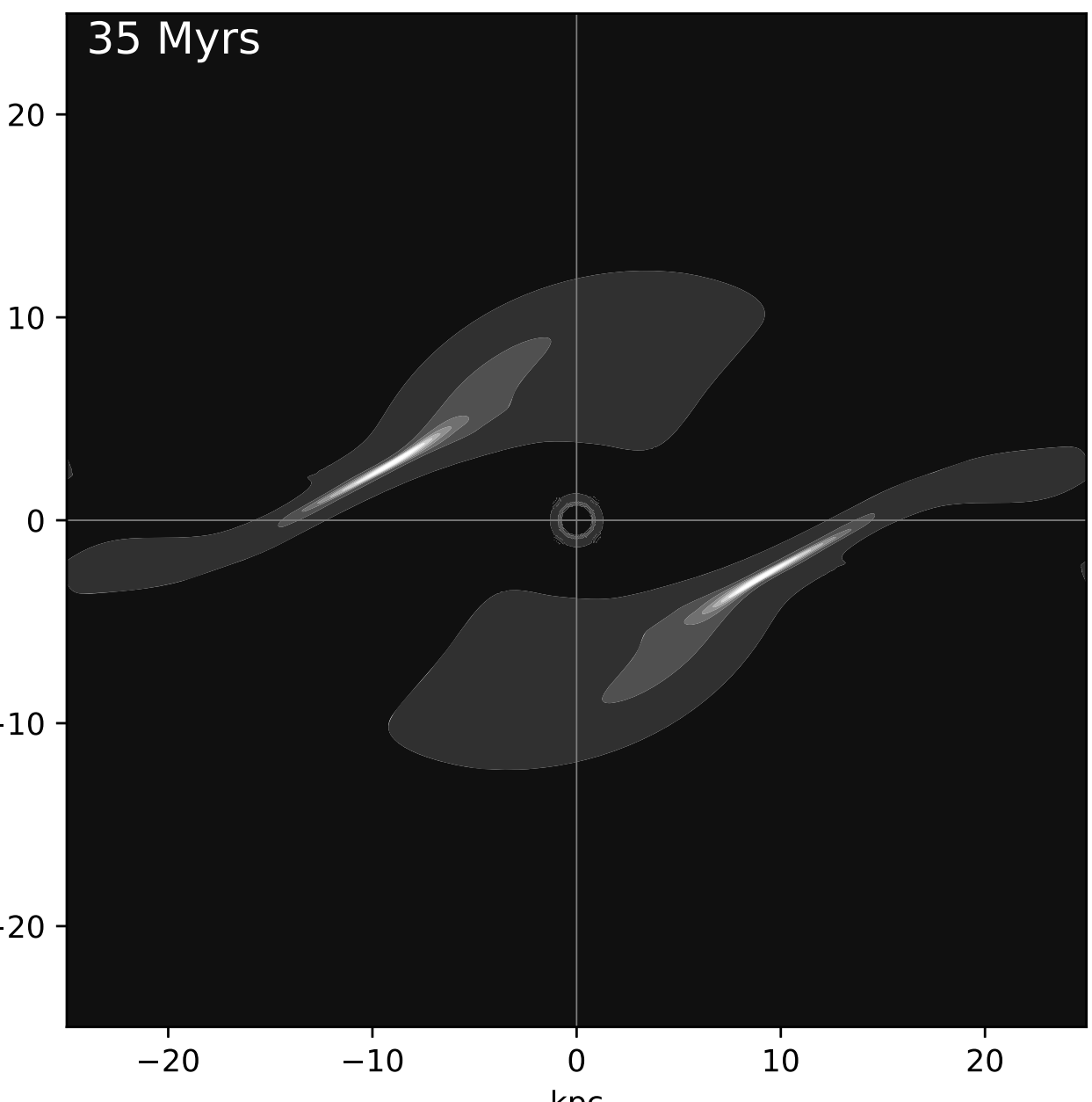

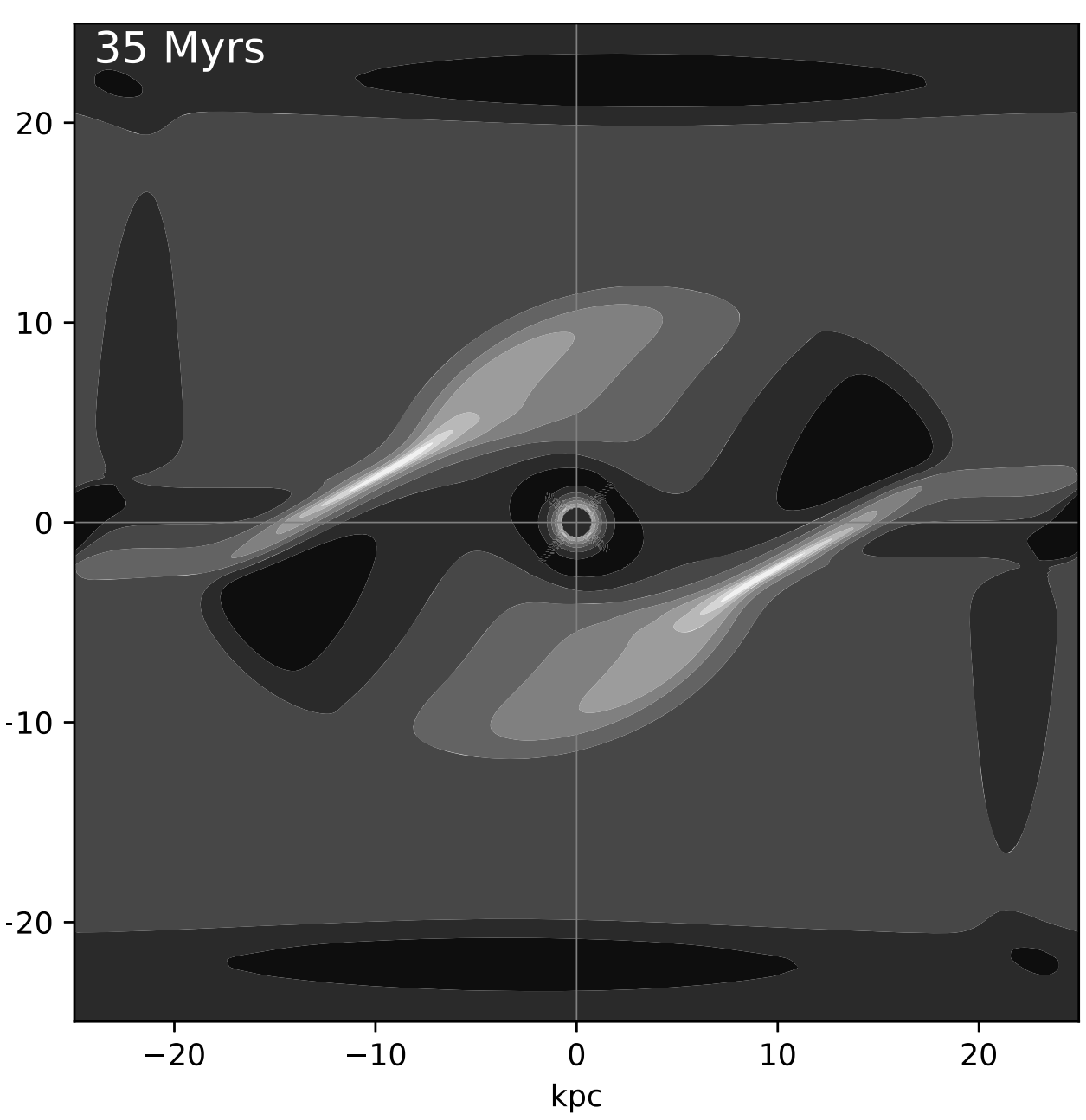

**Figure 8:** Contour plots for pressure/density 35 Myrs from initialisation, found by solution of the Navier-Stokes equation, showing the formation of a bisymmetric spiral density distribution. The upper plot has a linear scale between contours. The lower plot clips peak density and uses a log scale.

To better show the pressure/density distribution, within the vortex, I clipped pressures below the peak value and displayed contours using a log scale (figure 8 bottom). This clarifies the bisymmetric form and emerging spiral structure in pressure/density within the vortex.

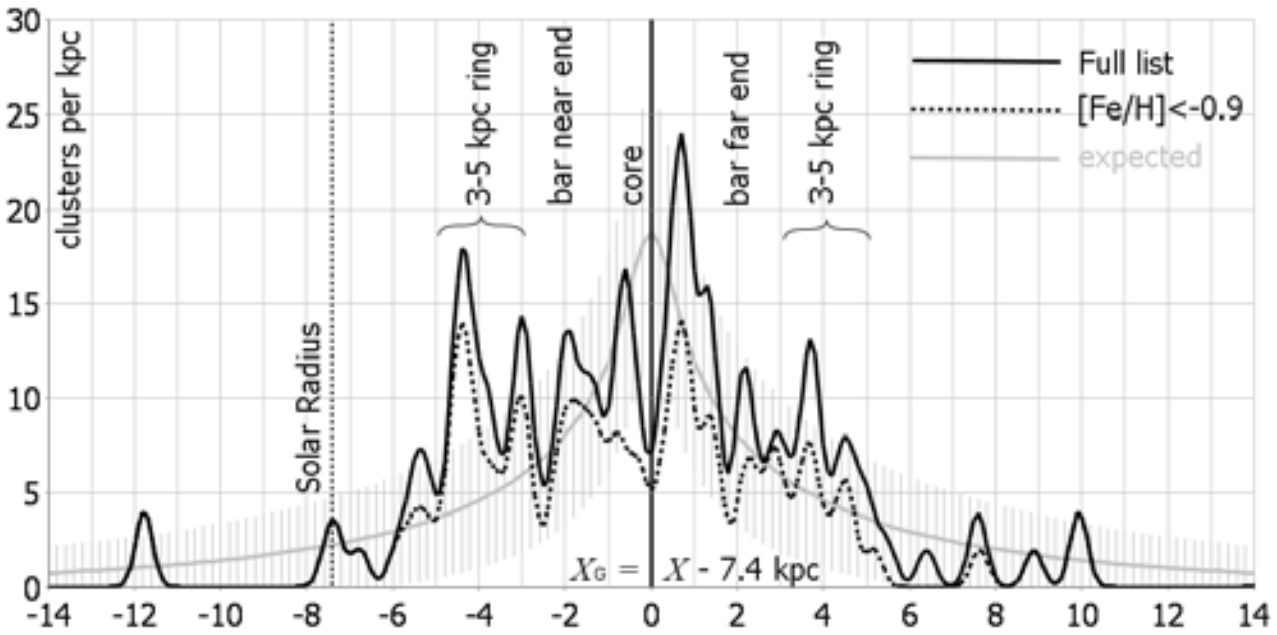

**Figure 9:** Distribution of cluster positions on an axis from the Sun to the Galactic centre for the full database, and for clusters with metallicities [Fe/H] < –0.9 (dotted). The scale has been shifted by 7.4 kpc, to show the Galactic centre at the origin. The expected distribution, projected onto the *X*-axis, for a spherically symmetric exponential probability distribution function with scale length 12 kpc (section 4) is shown in grey with error bars at intervals of the smoothing parameter (0.2 kpc).

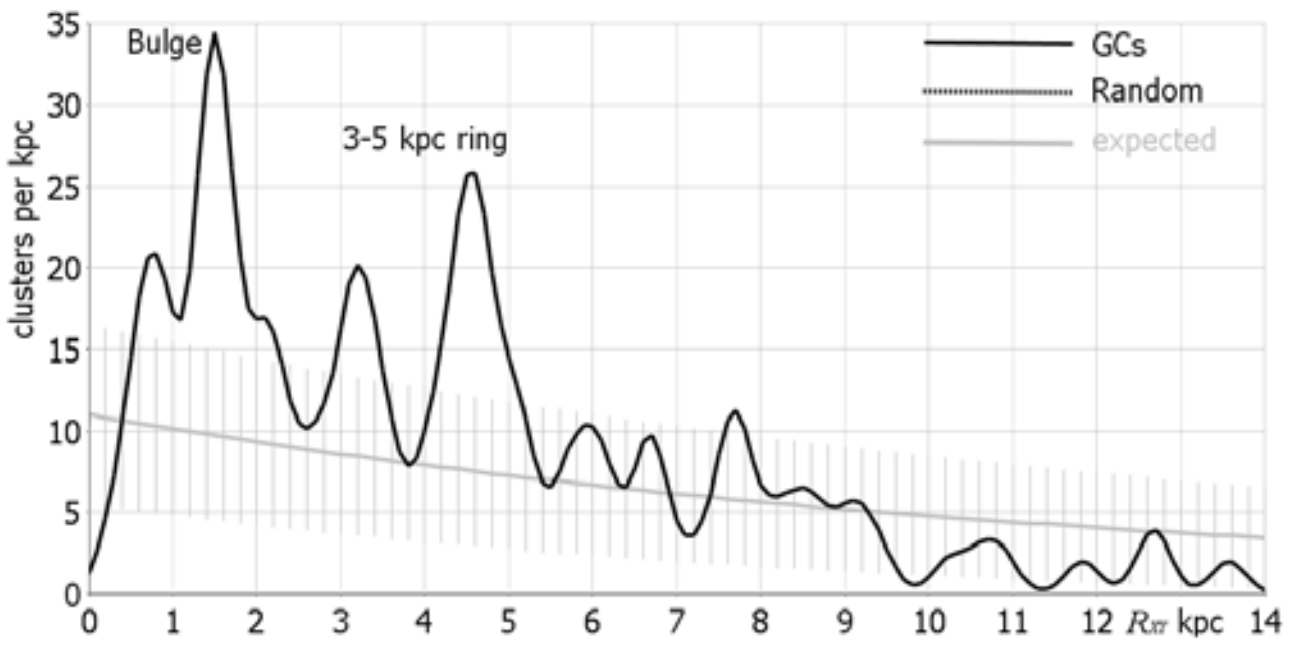

**Figure 10:** Distribution of cluster Galactocentric radial positions in cylindrical coordinates (black). A spherically symmetric exponential distribution with scale length 12 kpc is shown in grey with error bars at intervals of the smoothing parameter (0.2 kpc).

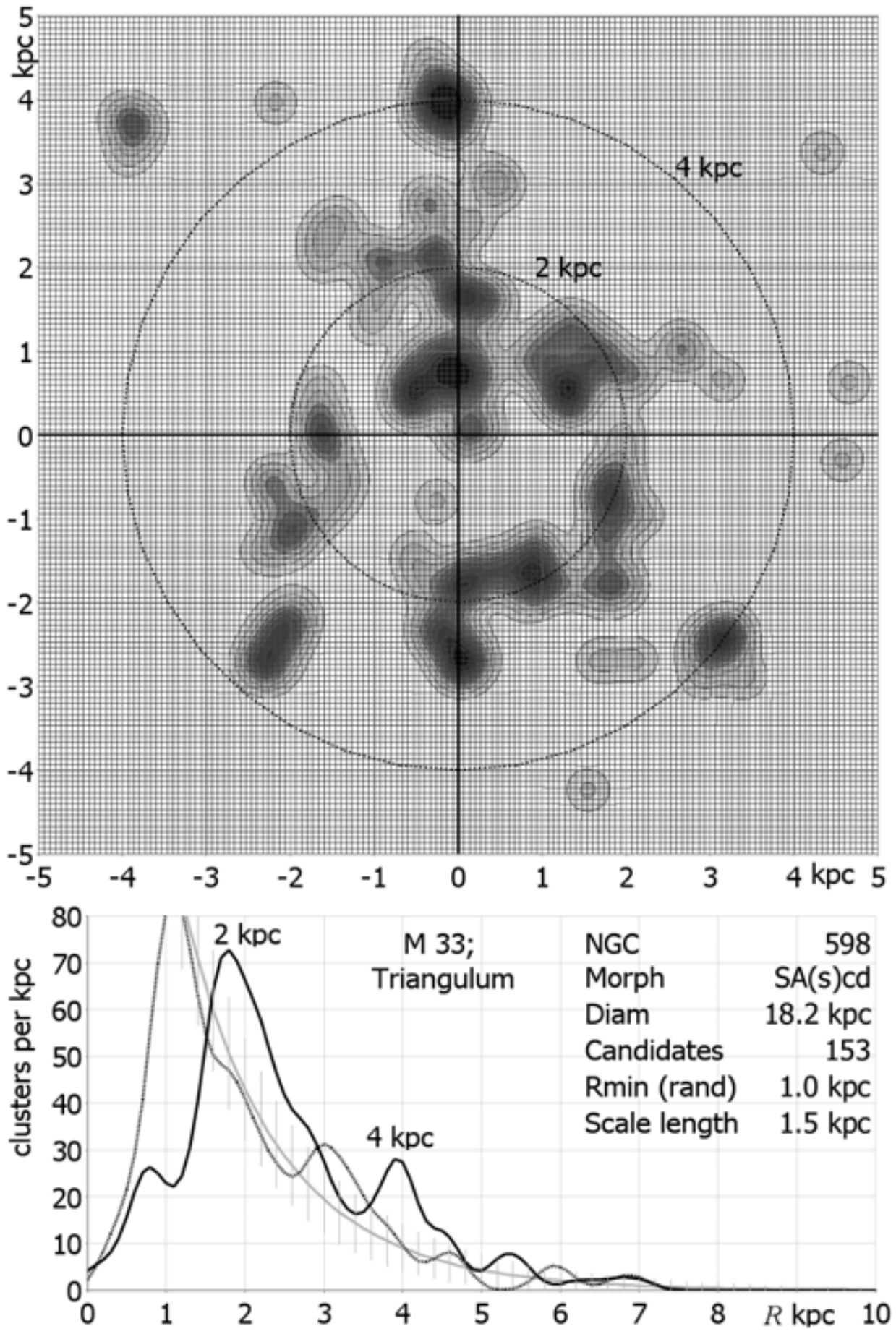

**Figure 11:** Distributions of globular clusters in M33. A projected exponential distribution is shown with error bars in grey. A random exponential distributions is shown dotted. Minimum radius 1 kpc and scale length 1.5 kpc is used (Vizier J/A+A/366/498).

In practice, galaxies grow over time from accretion of gas from the intergalactic medium. As observed in the Hubble Deep Field early galaxies were small. Simulations have shown that gas accretion is the primary process of galactic growth (e.g. Keres et al. 2005, Benson & Bower 2011). Direct evidence for gas accretion has been observed (Lehner et al. 2013, Crighton et al. 2013). The simulations performed here also show that gas accretion makes an essential contribution to bisymmetric spiral structure.

The radial distance of the high pressure regions from the central position depends only on the initial size of the vortex chosen for the simulation. The existence of these high pressure regions during early galaxy formation is confirmed by the observed distribution of globular clusters in the Milky Way and other galaxies and by the observation of multiple generations of early star formation in globular clusters (Gratton, Carretta & Bragaglia 2012 and references therein). After a globular cluster forms within one of the dense gaseous regions it will orbit the galaxy, triggering a new generation of star formation when it reenters a dense region. Francis and Anderson (2014) observed a significant overdensity of globular clusters at a radius 3-5 kpc from the galactic centre (figure 9). The overdensity is more clearly seen in galactocentric radial coordinates (figure 10). Similar overdensities are seen in M33 (figure 11) and M31 (figure 12). Since globular clusters lie on highly eccentric orbits and spend most time near apocentre, the radius of the overdensity may be taken as a measure of the radius at which the globular clusters formed.

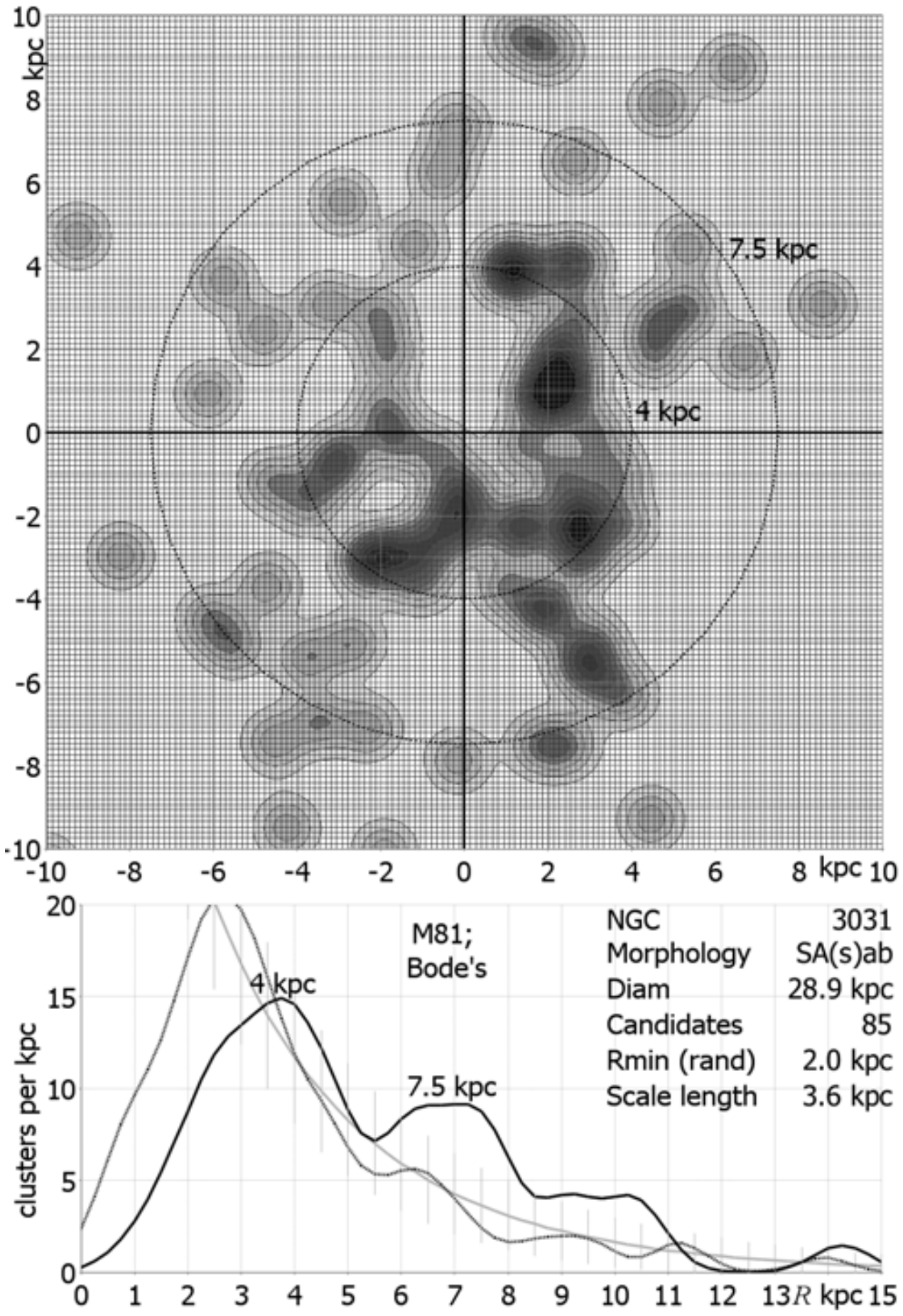

**Figure 12:** Distributions of globular clusters in M81. A projected exponential distributions is shown with error bars in grey. A random exponential distributions is shown dotted. Minimum radius 2 kpc and scale length 3.6 is used (Vizier J/AJ/142/183).

## 7 Gas motions in a spiral potential

I also modelled gas flow in a spiral potential, using an initial condition of circular motion and uniform density/pressure between radius 2 kpc and radius 16 kpc. Gravitational potential was determined by a central mass of $3.5{\times}10^{11}$ solar masses (representing net mass within 2 kpc of the centre) and spiral arms of radius 15kpc and $6.5{\times}10^{11}$ solar masses, pitch angle 11°, exponential mass distribution with scale length 8, and width 2 kpc determined by parameter $a$ in (8.2). Initial velocity was set with an exponential distribution with scale length 32 kpc and the boundary conditions at radii 2 kpc and 16 kpc were fixed such that pressure at these radii did not become extreme. This is done so that the model breaks down at turbulent star formation regions within the spiral arm, not at the boundary of the simulation.

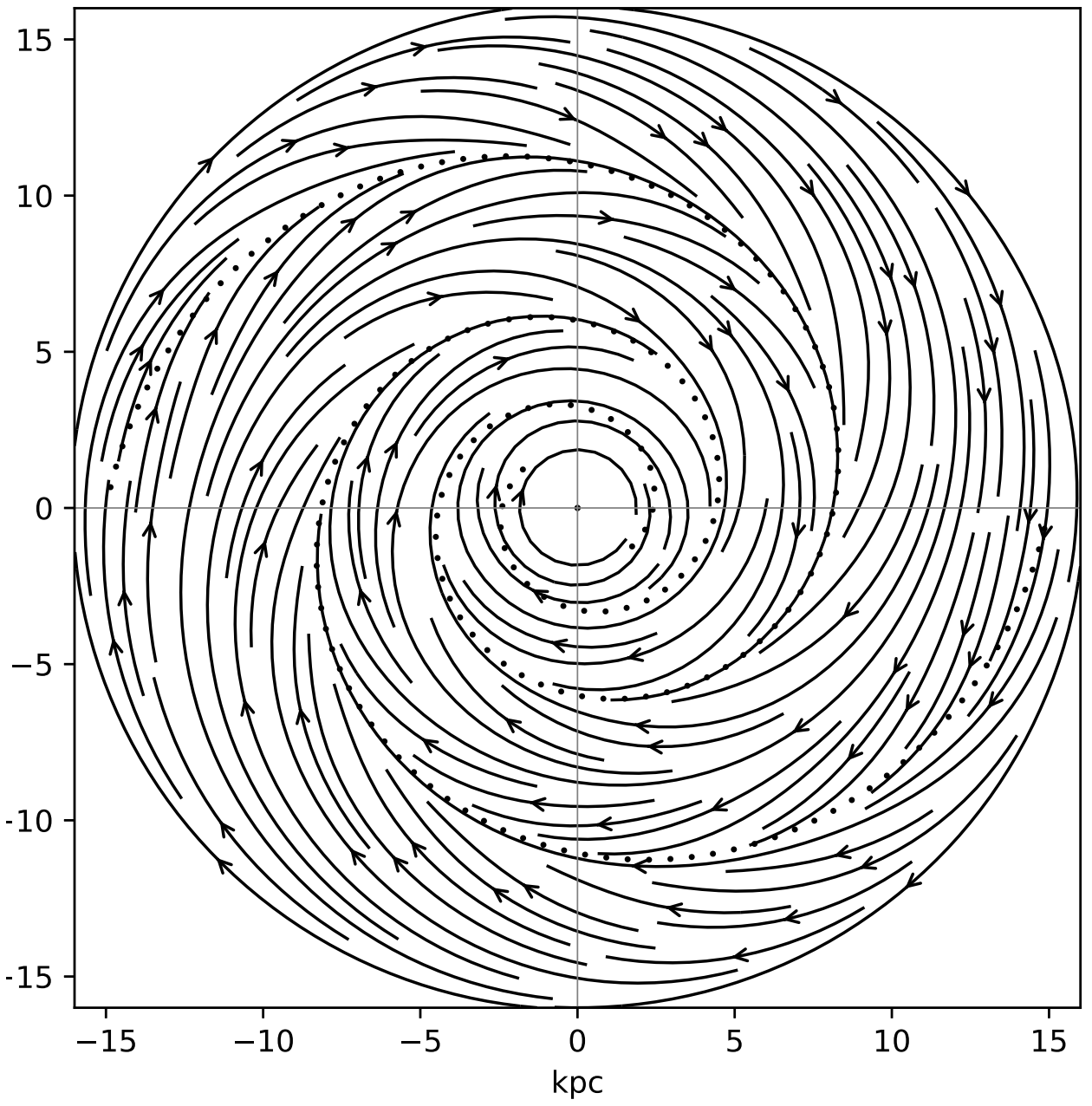

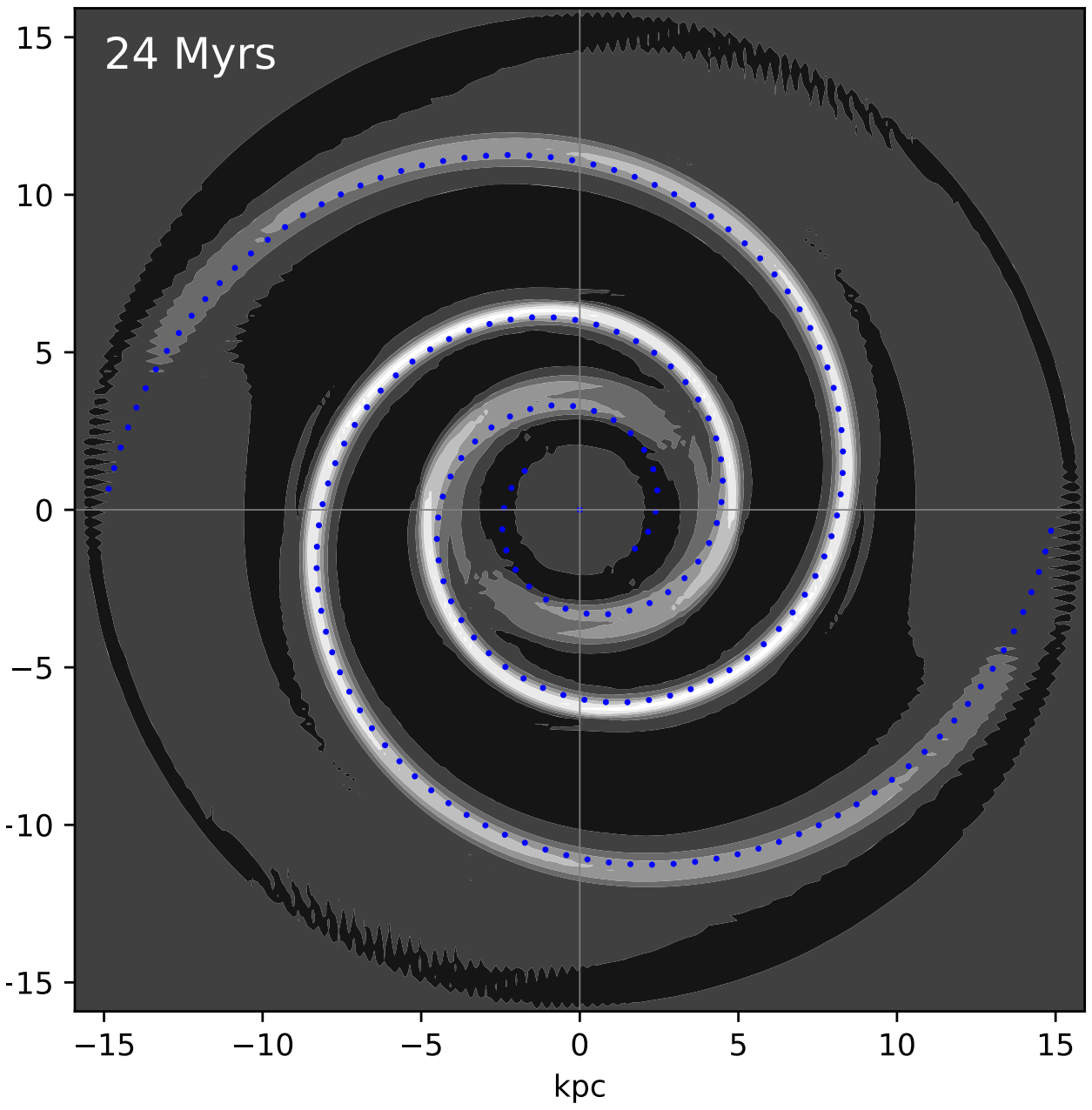

**Figure 13:** Digital solution of the Navier-Stokes equation in a spiral potential 24 Myrs after initialisation using circular motion and uniform pressure. Boundaries are set at radii 2 kpc and 16 kpc from the centre.

The model ran for 24 Myrs, at which point the velocity and pressure distributions are as shown in figure 13. Gas is drawn into the arms from both sides, creating regions of high density/pressure where turbulence arises and star formation will be triggered along the arms.

## 8 Potential in a spiral galaxy

For a conservative vector force field, $\mathbf{F}(\mathbf{x})$, such as gravity, we can define a scalar potential field, $V(\mathbf{x})$ such that

$$\mathbf{F} = -\nabla V, \tag{8.1}$$

where the differential operator, $\nabla$, gives the gradient of $V$ in the direction of the greatest slope.

For constant $a$, define a function

$$K(u) = \begin{cases} -\dfrac{G}{u} & \text{for } u > a \\ -\dfrac{G}{a} & \text{for } u \le a \end{cases}, \tag{8.2}$$

where $G$ is Newton's universal constant of gravitation. The potential outside a spherical gravitating body of mass $M$, radius $a$, and with position vector $\mathbf{r}$ is described by a potential well with the form

$$V(\mathbf{x}) = MK(|\mathbf{x} - \mathbf{r}|). \tag{8.3}$$

For a system, such as a galaxy, with many particles (stars) of mass $M_i$ at positions $\mathbf{r}_i$, the potential can be summed over the particles,

$$V(\mathbf{x}) = -\sum_i GM_i K(|\mathbf{x} - \mathbf{r}_i|), \tag{8.4}$$

where, for convenience, the actual radius of each star is replaced with a constant radius $a$, representing a maximum stellar radius. Equation (8.4) is exact at a distance greater than $a$ from any star. Since (8.2) is proportional to a kernel function it is immediate that the potential (8.4) is proportional to the kernel density estimate of the mass distribution. That is to say, potential is proportional to smoothed density in the region where there is mass (the inverse law dominates outside of this region). For an N-body simulation the same argument applies, but particle mass may be taken as constant and representative of a large number of stars with a uniform density distribution within in a sphere of radius $a$.

A plot of the gravitational potential for a spiral galaxy (vertical axis) against the plane of the disc for a typical spiral galaxy in which mass is concentrated towards the centre and in the arms is shown in figure 14. Neglecting perturbations due to individual stars, the description of stellar orbits in such a potential is analogous to the trajectories of particles in a frictionless spiral-grooved funnel in a uniform gravitational field for which potential is directly proportional to height. Different mass distributions will affect the size and shape of the funnel and of the grooves, but stellar motions are qualitatively similar for a wide range of spiral potentials.

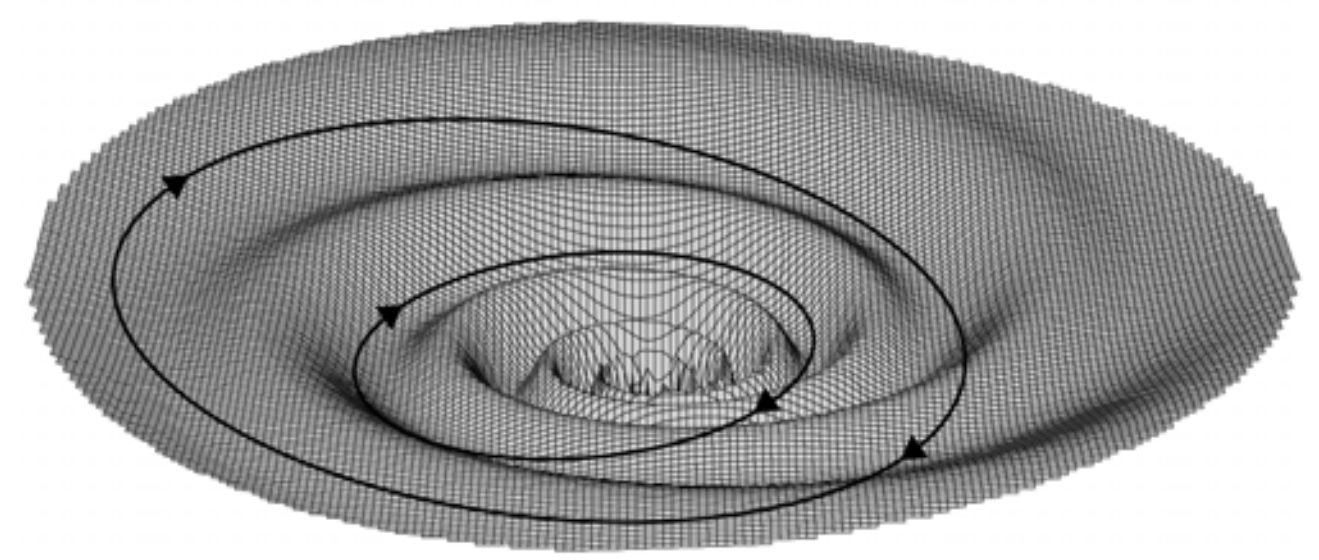

**Figure 14:** The gravitational potential of a bisymmetric spiral galaxy with an exponential mass distribution plotted on a vertical axis against the galactic disc on a horizontal plane. Potential closely follows the mass distribution. The alignment of osculating elliptical orbits with troughs in the potential is shown. True orbits precess in both directions while remaining aligned with the arms.

A particle at the highest point of its path, where it is moving least quickly, will tend to fall into a groove and then follow the groove downwards, picking up speed as it goes. Thus orbits follow the arms on the inward part of the motion, as is seen in the velocity distribution. Eventually, the particle gains enough momentum to jump free of its groove. It crosses over the next-highest groove (for a bisymmetric spiral), then falls back to a higher point in its original groove. The crossing point of the other arm takes place at a precise part of the orbit, and accounts for the Hyades stream in the Solar neighbourhood.

Orbital alignments with spiral arms can then be understood using a model constructed by repeatedly enlarging an ellipse (representing the osculating orbit) by a constant factor, $k$, centred at the focus and rotating it by a constant angle, $\tau$, with each enlargement (figure 15). The pitch angle of the spiral depends on $k$ and $\tau$, not on the eccentricity of the ellipse; for a given pitch angle, ellipses with a range of eccentricities can be fit-ted to the spiral, depending on the desired width of the arm and proportion of the circumference of each ellipse embedded within it. Higher eccentricity orbits fit spirals with higher pitch angles. Thus stars move along an arm on the inward part of their orbits, leaving the arm soon after pericentre, crossing the other arm on the outward part and rejoining the original arm before apocentre.

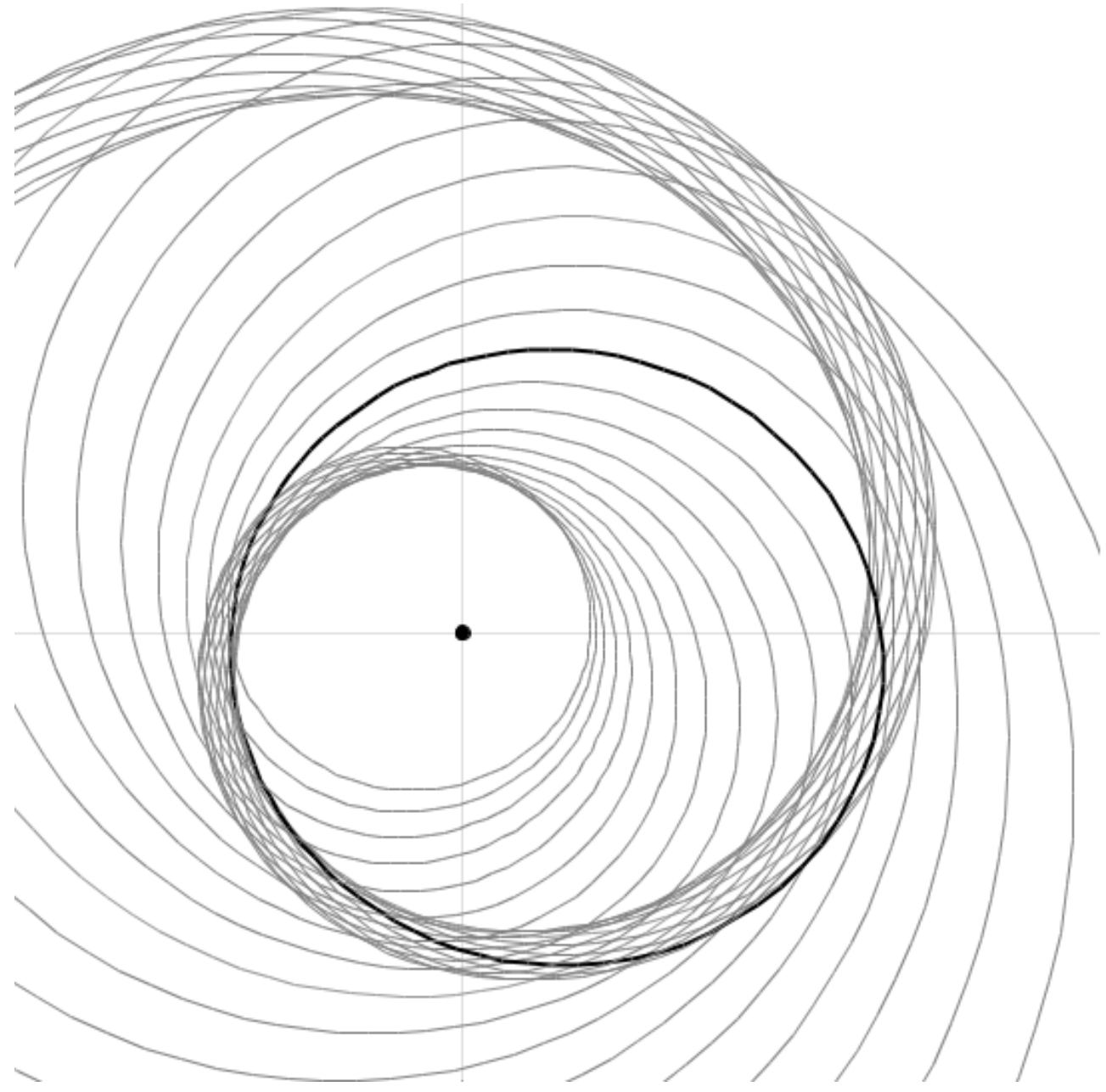

**Figure 15:** An equiangular spiral with a pitch angle of 11°, constructed by repeatedly enlarging an ellipse with eccentricity 0.3 by a factor 1.05 and rotating it through 15° with each enlargement. Ellipses with eccentricity greater than about 0.25 have more than half their circumference within the spiral region. Ellipses with eccentricity greater than about 0.35 produce probably too broad a structure to model a spiral arm with this pitch angle, but give a good fit for spirals with greater pitch angle. Lower eccentricity ellipses produce a narrower spiral structure and/or a fit with spirals of lower pitch angle.

The increase in stellar density in the arms is due to the paths followed by stars. The tendency for stars to follow the arms reinforces the gravitational potential of the arm. Thus, the potential is a consequence of the mass distribution and the mass in the arms is reinforced by the potential. Such a model is insensitive to the shape of the funnel and can account for the observed frequency of spirals in galaxies with a wide range of sizes and mass distributions.

## 9 N-body simulation

I constructed an N-body simulation. The use of a standard desktop PC, running a 4.2 GHz Intel i7 CPU limited the maximum practicable number of stars to 50 000 and the amount of testing of galactic parameters, but it was possible to establish durable spirals of different types for a wide range of parameters (It is hoped to port the code to a graphics card which would be expected to allow an order of magnitude increase in the number of stars, as well as testing a greater range of parameters).

The model calculates in the Newtonian formulation using Cartesian coordinates in three dimensions. Although spiral structure is essentially two dimensional, modelling in three dimensions adds only about 5% overhead, and has the benefit of reducing scattering by near collisions, this being a principle cause of degradation of spiral structure.

For definiteness, I set disc mass equal to $1.0\times10^{11}$ solar masses in a galactic radius 15 kpc, with a Gaussian cross section with standard deviation 0.2 kpc. I modeled spiral arms in the range 2 to 15 kpc using a random exponential distribution with scale length 8 for orbits at apocentre. Actual scale length is less, because stars are not all at apocentre.

As the model does not solve the Navier-Stokes equations or describe star formation, I adjusted galactic potential for gas and dust. As seen in section 7, gas is expected to follow the arms. Greater quantities of gas can be expected in the outer regions, because in the central region gas is depleted by star formation. I set a linear relation such that gas density increases in proportion to the distance from the galactic centre. The total amount of gas used was respectively 10%, 15% and 20% of arm mass (and a lower proportion of total galactic mass). These proportions may underestimate gas and dust, because most gas is expected at the outer edge of the galaxy where cold gas is harder to detect. They do not include a contribution from cold dark matter.

Since we are interested in the behaviour of stars in the arms, I did not attempt to accurately model the central region where velocities and accelerations are large. The bulge and core were treated as a single body at the galactic centre. Modelling parameters were chosen after testing models with 5 000, 10 000 and 20 000 stars, from which it was ascertained that the qualitative behaviour of the model is not sensitive to any particular choice of parameters chosen within reason. In models with fewer stars spiral structure decays more quickly, due mainly to hard scattering and random clumping of the stellar distribution which introduces irregularities in spiral potential. However, it is perfectly possible to model an enduring spiral with only 5 000 stars provided that initial conditions are well chosen.

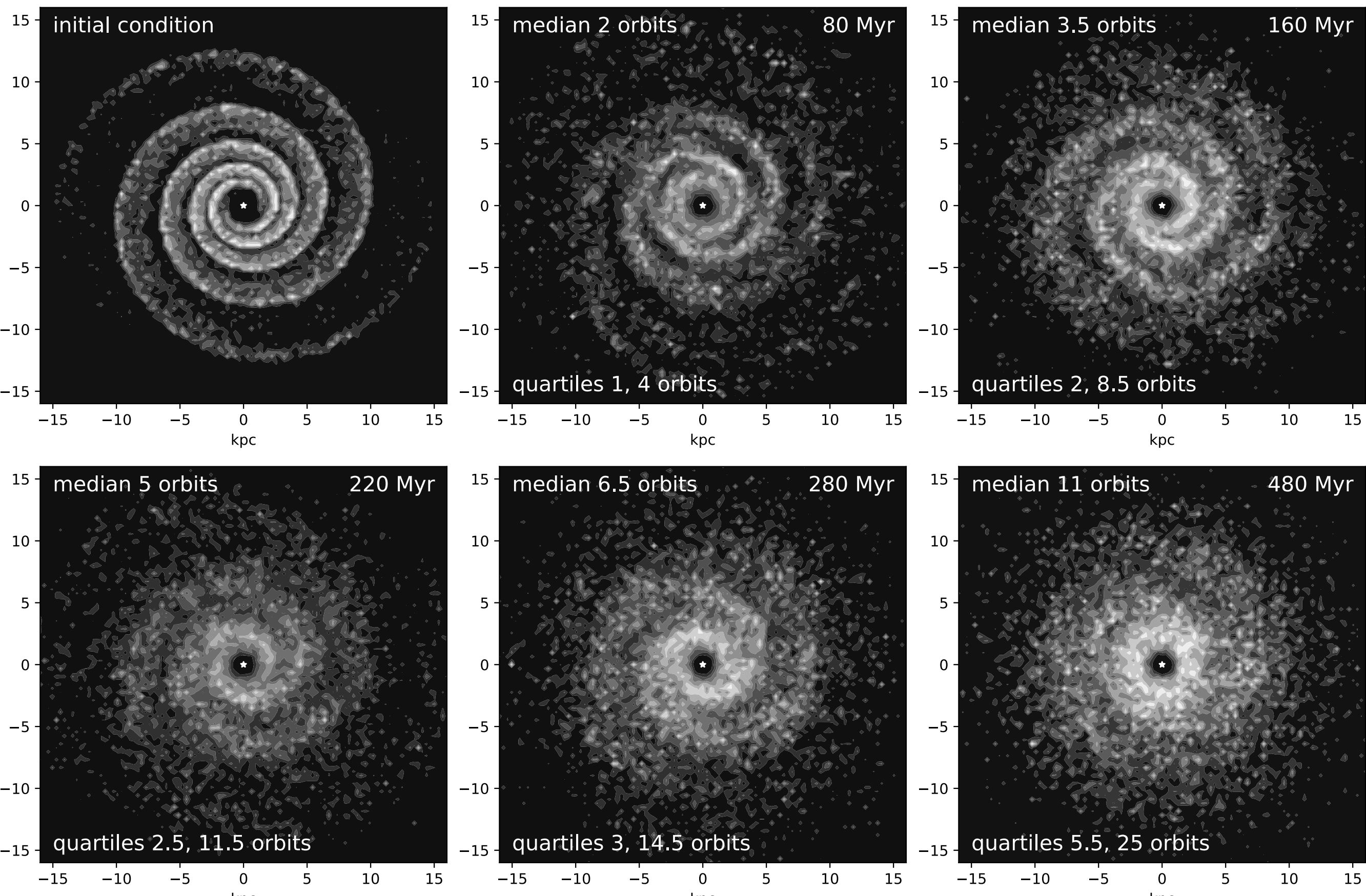

**Figure 16:** Contour plots of stellar density in the evolution of a model Sa spiral galaxy with only 50 000 stars. Mean orbital eccentricity decays from 0.30 to 0.29. Eccentricity dispersion increases from 0.07 to 0.14. Half orbits are counted when a star crosses the *x*-axis.

Time steps were chosen to ensure negligible precession for all orbits under the gravity of only the central mass. This requires for an Sa galaxy with orbital eccentricities 0.3 a step of $1\times10^5$ years, a minimum of 130 steps per orbit. For an Sb galaxy with orbital eccentricities 0.5, it is $7\times10^4$ years, or at least 150 steps per orbit. For an Sc galaxy with orbital eccentricities 0.7, the step is $3\times10^4$ years, or at least 240 steps per orbit. Since the arms and gas lie outside the core, and individual stars exert a low acceleration for sufficient $N$ and appropriate smoothing parameter, $a$, they do not necessitate a reduction in the time step.

Noting that the smoothing parameter $a$ in (8.2) is expected to be inversely proportional to the square root of the number, $N$, of stars I established a suitable choice,

$$a = \sqrt{\frac{50}{N}}\ \text{kpc}, \tag{9.1}$$

by minimising the increase in eccentricity dispersion (too small a value of $a$ introduces hard scattering due to near approaches, while too large a value reduces the depth of the trough in the potential responsible for orbital alignments).

The choice of initial conditions for the simulation is the most important single factor governing whether a spiral has an enduring structure. This is expected since damping forces are negligible. As stellar winds travel away from stars faster than the speed of sound, the maxi-mum speed at which a force can be transmitted through a gas or a plasma, no drag force can be transmitted to the star from the interstellar medium. The simulations confirm that gravitational damping is negligible as orbital eccentricities change slowly during evolution.

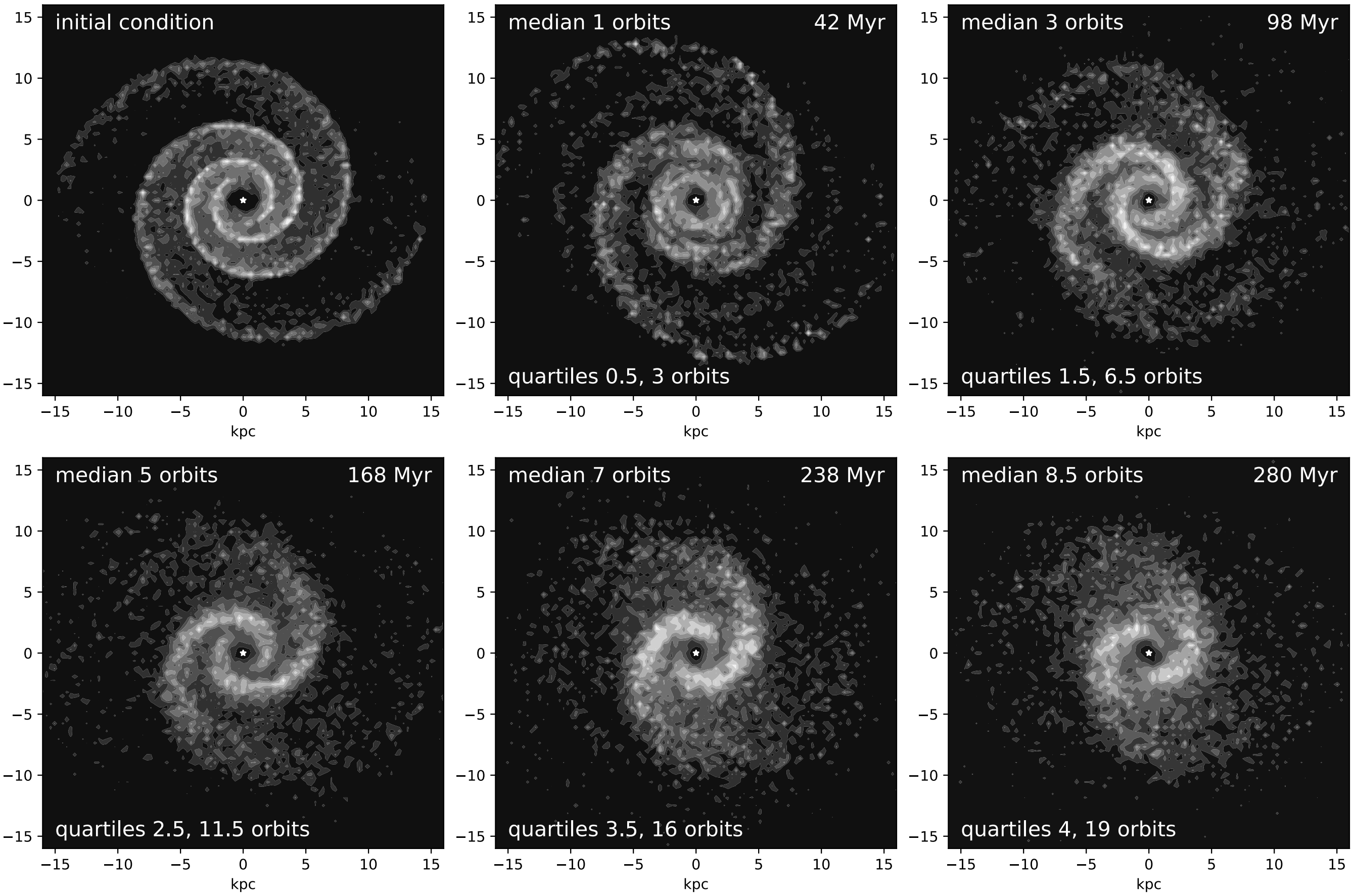

**Figure 17:** Evolution of an Sb galaxy with only 50 000 stars. Mean orbital eccentricity decreases from 0.49 to 0.47 during the course of the simulation. Eccentricity dispersion increases from 0.07 to 0.19.

In the absence of significant damping, it is necessary that initial conditions for simulations are close to dynamically near-stable conditions for spiral structure. In practice, for the models tested, spiral structure decays slowly over time. Increasing the number of stars increases the duration of the spiral but it appears that initial conditions have a greater importance, and that for a stable model it will be necessary to model gas motions and star formation to arrest the decay of the spiral from the outer edge.

The initial condition for the simulation uses a random distribution of stars at arbitrary position on osculating elliptical orbits on with eccentricities close to 0.3, 0.5 and 0.7, designated by Hubble types Sa, Sb and Sc respectively. Initial orbits (based on a calculation of contained mass) were normally scattered with an angular standard deviation of 0.1 radians either side of logarithmic bisymmetric spirals with pitch angles 8°, 11° and 30°, for types Sa, Sb and Sc. A further correction to velocities was made to take into account that stars move faster when they fall into the potential trough due to spiral structure.

It was immediately clear that these velocities are too low as the galactic radius shrinks significantly during the initial evolution. To stabilise the initial behaviour of the galactic radius, velocities for gasless Sa and Sb galaxies must be increased by an amount equivalent to an effective dark matter content, after inclusion of gas and dust, of about 30% of arm mass. Rotation of the spiral pattern is reduced and spiral structure also lasts longer. This is not an indication of dark matter but is a direct consequence the gradient of spiral potential due to conventional matter in Newtonian gravity. It can be understood because, as seen in figure 15, a star lies on the outer part of the spiral arm for a substantial arc during the outermost section of its orbit, where gravitational attraction due to the arm dominates (figure 14). Since force is proportional to the gradient of the potential, inward radial force is greater on the outer wall of the potential trough than would be the case in the absence of spiral potential. The increase in inward acceleration during the outer part of the orbit requires that stars orbit more quickly than they would in an axisymmetric potential. The effect is more complicated for open-armed Sc galaxies because with greater pitch angle the radial component of inward acceleration is less and because the trough wall also imparts a non-negligible retardation force. The simulation indicates a small reduction in effective mass for Sc galaxies.

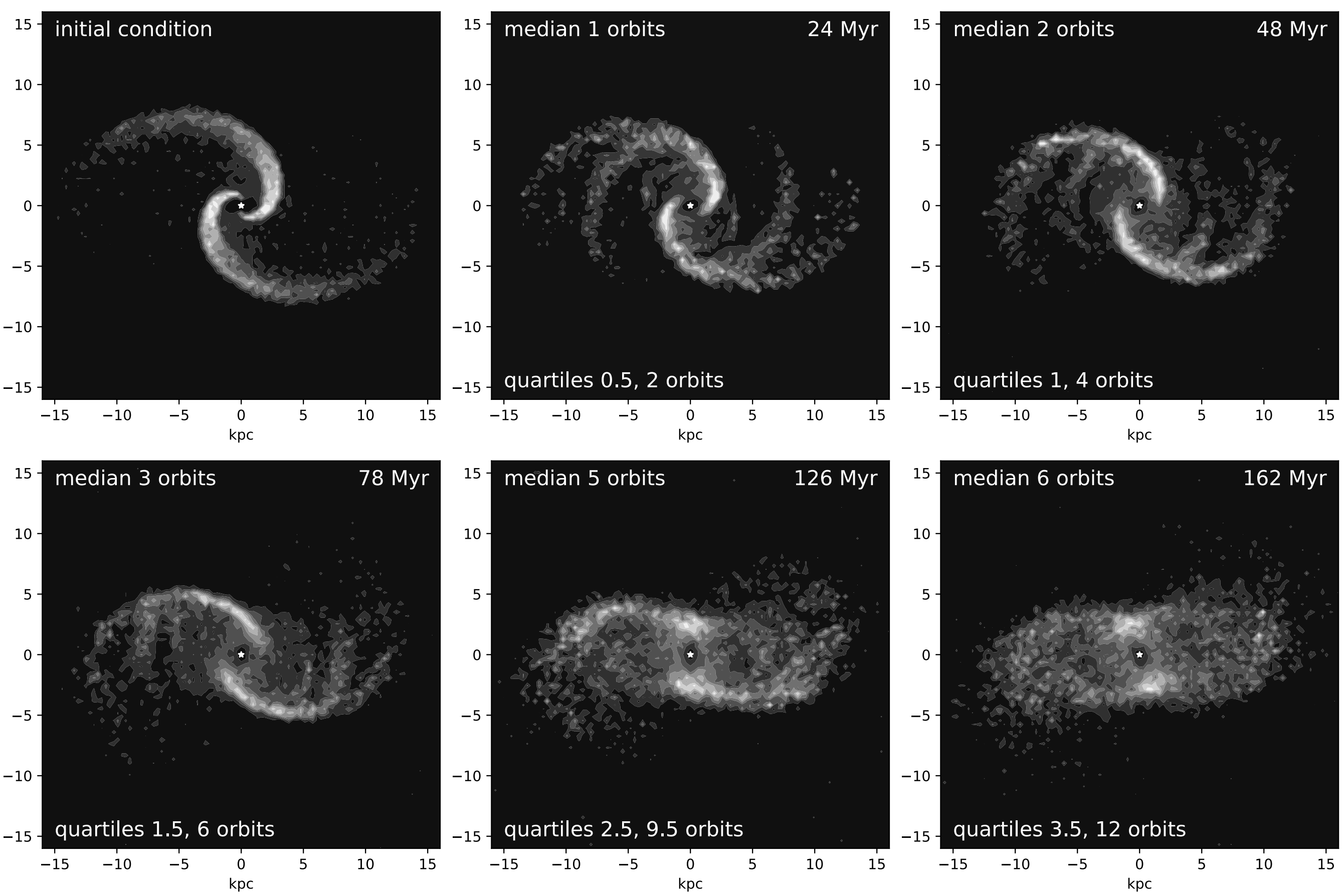

**Figure 18:** Evolution of an Sc galaxy with only 50 000 stars. The appearance of additional arms is due to difficulty in establishing a realistic initial condition for a stable solution. The formation of interlocking rings may be realistic. Mean orbital eccentricity falls from 0.69 to 0.65 during the course of the simulation. Eccentricity dispersion increases from 0.06 to 0.23.

I ran the model with 50 000 stars, plotting the results at various time slices for Sa, Sb and Sc galaxies as shown in figure 16, figure 17 and figure 18 respectively. In each case spiral structure endures for a number of stellar orbits and degrades slowly. The timescale for evolution is conveniently measured by the median number of orbits of an arm star, counted as a half-orbit each time a star crosses the $x$-axis. A major cause of decay is due to scattering. This is shown because in each model there is little change to the mean orbital eccentricity during the period of the simulation, but dispersion in eccentricity increases, while the range increases substantially. A significant increase in durability was found by increasing the number of stars from 25 000 to 50 000. It seems likely that significantly greater durability can be obtained by increasing the number of stars to reduce scattering. This will require more processing power.

The spirals appear to shrink over time. This is primarily because stars reaching the outer edge of the galaxy do not accurately rejoin the spiral arm. It is to be expected that in a real galaxy gas motions and star formation replenish the outer arm structure through accretion, particular from regions of high gas density identified in figure 8.

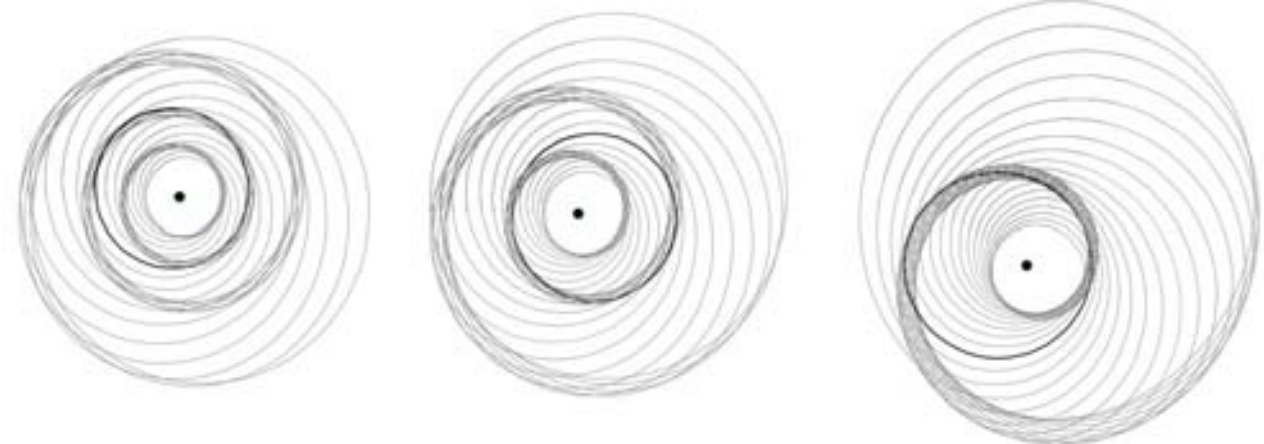

**Figure 19:** Observations of stellar motions in the Solar neighbourhood have shown that orbits are aligned with spiral arms, and follow the trough in gravitational potential created by a spiral arm along the inward parts of their orbits. Galaxies in which orbital eccentricities are greater generate spirals with greater pitch angle.

The decay of the Sc galaxy of figure 18 shows signs of evolving towards interlinked rings. This form may not be due to modelling errors. This pattern was predicted by Francis & Anderson (2009) in consequence of arm densities becoming dense toward the galactic centre and is observed, for example in UGC12646, ESO 325-28, NGC 2665 and NGC 619.

## 10 Morphology

The morphology of a spiral galaxy is determined by the properties of the vortex from which that galaxy formed. Gas in a vortex moves in approximately circular motion. Because gas is an elastic medium, waves in density can form along the angular direction such that an integer multiple of wavelength is equal to circumference, generating the familiar spiral form of a vortex. The boundary condition is a flow parallel to one axis (the horizontal axis in figure 6), and no flow across the perpendicular axis. This creates a forcing term in the Navier-Stokes equation (transmitted radially by viscosity) with two cycles per orbit. Thus gas flows generate a bisymmetric spiral vortex, as seen in section 6, triggering star formation in the regions of highest density.

In a vortex, gas flow is near circular. For typical spiral galaxies, initial star formation is at apocentre. As described by Sofue and Rubin (2001), spiral galaxies with tightly coiled arms (Sa) generally rotate more rapidly than more open spirals (Sb & Sc). The rate of rotation at time of formation determines orbital eccentricity. Higher orbital eccentricities result from slower rotation rates, leading to the formation of spiral arms with greater pitch angles (figure 19). There is also a correlation with mass. Lower mass galaxies tend to have more open spirals, because smaller vortices form in oblique collisions between more slowly moving gas clouds and in secondary vortices where rotation rates are reduced.This is also predicted by the model of spiral structure (after star formation) described by Francis & Anderson (2009, 2012). Stars form in regions of high pressure created between colliding ingoing and outgoing gas flows. The result of collision is generally low radial velocity, so that stars form near to orbital extrema.

If the infalling cloud was not rotating, a boxy elliptical galaxy was formed with no net rotation. With some rotation of the infalling gas, a rotating, discy, elliptical galaxy was formed. At greater rotation, stars were formed at apocentre in high eccentricity orbits, leading to an Sc spiral galaxy with open arms. With still more rotation, eccentricities are further reduced, leading to a tighter spiral structure. When the rotation of infalling gas was great enough for near circular motions, a lenticular galaxy was formed, in keeping with observations that orbits in lenticular galaxies are near circular. Still faster rotation may lead once again to spiral structure, but if orbits are for stars formed at pericentre, stars will move outward from the radius at which they form, explaining the lower density and greater spread of low surface brightness galaxies. In comparison to typical spirals, low surface brightness galaxies are usually more isolated. This agrees with the expectation that greater rates of gas flow should be found in more rarefied regions where damping processes have been less.

In order of increasing rotation from left to right, this suggests the morphology shown in figure 20. This ordering is intended to emphasise the dynamical properties of gas clouds leading to different types of galaxies. Other orderings will result from a different choice of criteria. For example, star formation is dependent on collisions between gas clouds in spiral galaxies and is largely absent in elliptical and lenticular galaxies. Following Kormendy and Bender (1996) the subclassification of ellipticals is ignored, since it has been found to be largely dependent on the angle of inclination. Discy ellipticals, which are rotating, are distinguished from boxy ellipticals, which are not. The morphology reverses the order of spiral types and lenticular galaxies in the usual Hubble sequence, so as to show increasing rotation rates towards the right, in accordance with the observation reported by Sofue and Rubin and the model of spiral structure described by Francis & Anderson. Low surface brightness galaxies are shown on the far right, as their formation can be expected from the gravitational collapse of faster rotating vortices. Faster rates of initial rotation are predicted to produce more open low surface brightness spirals.

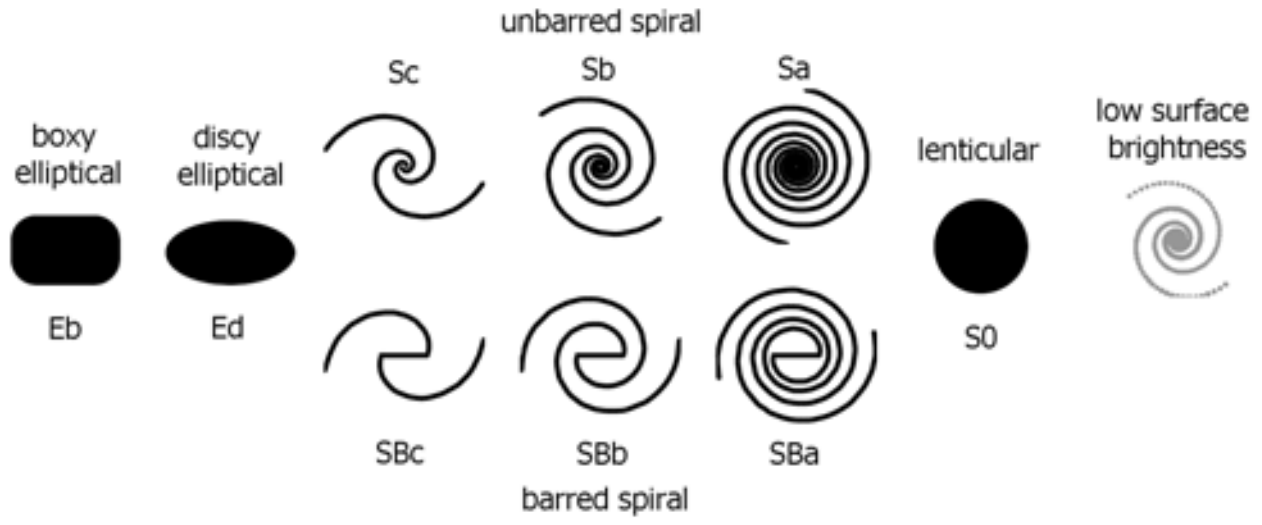

**Figure 20:** Revised galaxy morphology, with faster rotation at formation towards the right.

## 11 Rotation curves

In the inner region of a spiral galaxy, where gas is depleted due to star formation, stellar dynamics is governed by Newtonian gravity. In this region, as observed in section 9, stellar velocities are greater as a result of spiral potential. Gas motions are governed by the Navier-Stokes equation and there is a contribution due to pressure as well Newtonian gravity, but there is broad agreement between rotation curves in gas and in stars. As described by Sofue and Rubin, the flattening of rotation curves is mainly observed in gas in outer regions, outside the diameter of visible stars. In this region acceleration due to gravity is low and dynamics is governed by gas pressure. The boundary condition of a flat rotation curve requires that pressures equalise such that the net of gravity and pressure follows an inverse law. Thus, a typical spiral galaxy is effectively governed by the formula (up to the definition of constants) given by MOND, the phenomenological law proposed by Milgrom in 1983, which retains the inverse square law in the central region, but replaces it with an inverse law at greater radii. In the present proposal, modification of Newtonian dynamics is replaced by a term from the Navier-Stokes equation describing to gas pressure.

Hundreds of galaxies of many types, from typical spirals to dwarfs and low surface brightness galaxies have been fitted to MOND (see e.g. Sanders & McGaugh, 2002 for a review). Any further fitting here would thus be redundant. Significantly, MOND cannot model abnormalities like counter-rotating cores, asymmetric curves and steeply declining rotation curves in outer regions. Changes of flow over time and turbulence are expected to produce precisely these kinds of chaotic behaviours. Lower mass galaxies have less gravitational effect on the flow of surrounding gas. For this reason their *apparent* dark matter content is greater. Thus, it appears that no issue of dark matter arises with rotation curves. The curves show the motion of gas in the interstellar medium, and are determined mainly by large-scale, extragalactic processes, not by dynamical equilibrium with only the gravity of the galaxy itself.

## 12 Violation of the cosmological principle

A "buckyball" geometry perturbs a Friedmann cosmology, by concentrating mass in walls between cells (figure 21). A realistic model will have a more random structure than that shown. Space within each cell will be near flat because according to Newton's shell theorem, which may be held valid in approximation, the gravity of surrounding walls will cancel within a cell — the corollary is that voids form when matter falls away from low density regions, rather than being drawn towards dense cell walls. To a first approximation we would not detect the gravity of the largest walls. The apparent flatness of space found from the spectrum of cosmic background radiation is accounted because space between the greatest walls is expected to be near

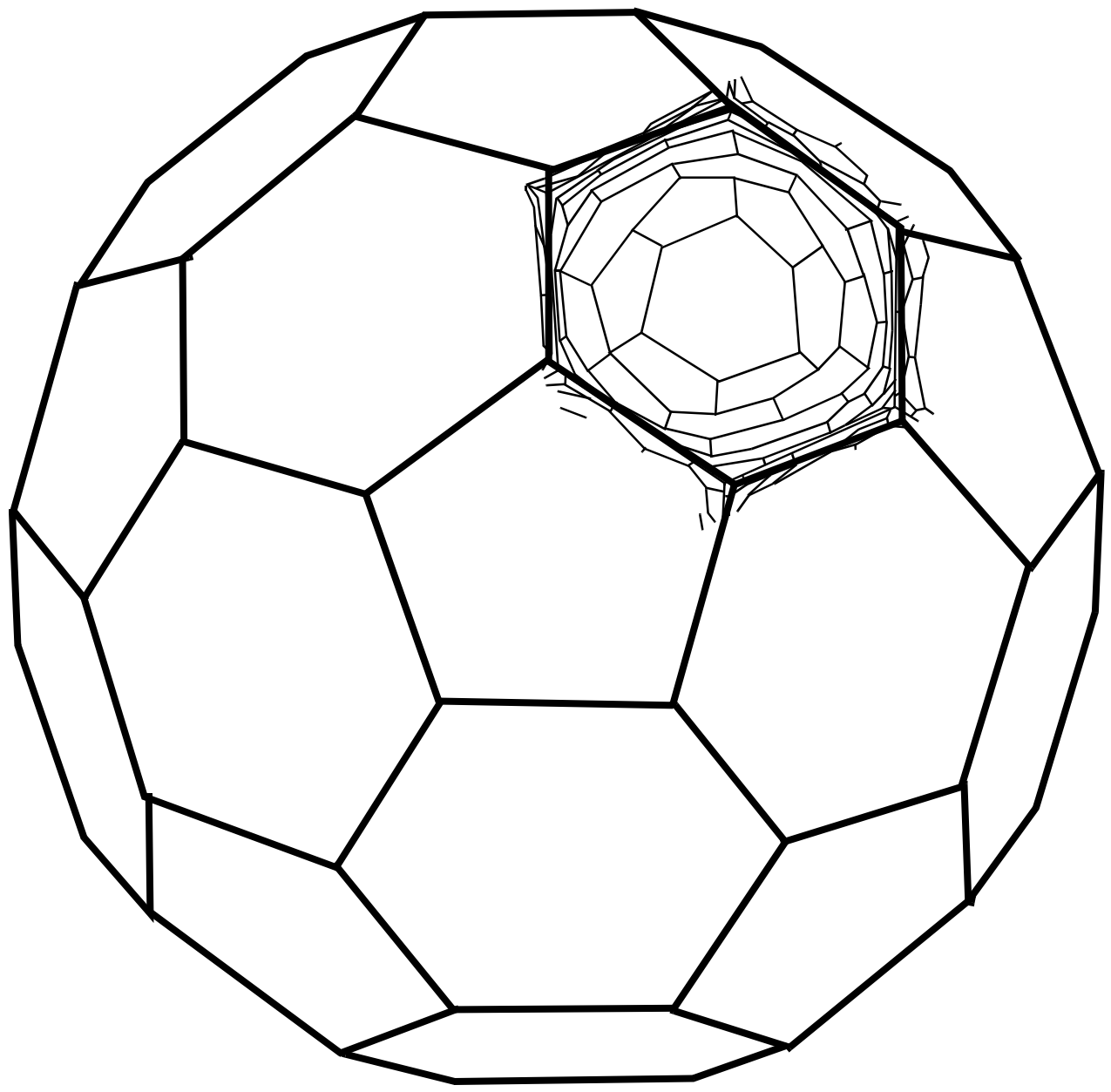

**Figure 21:** The geometry of a space-like slice through the universe at constant cosmic time may more closely resemble a buckyball than a space of constant curvature. The cosmological principle is obeyed in approximation on the scale of the universe as a whole, but curvature is concentrated in massive walls resulting from collisions in gas flows emanating from the Big Bang. Each face is tessellated, but the bulk of curvature is in the most massive walls. A realistic model would randomise the structure. The number of faces is unknown. Open as well as closed models are possible. The scale is such that the observable universe is contained within a single face, explaining the apparent flatness found in the analysis of the cosmic microwave background.

to flat, and because we only observe a part of one face of a "buckyball" universe.

The position of the most massive walls can be taken as approximately constant in comoving coordinates since they are formed from material infalling from cells on both sides. We can then write down a convenient form of metric of an expanding inhomogeneous cosmology using the Stephani metric, which makes use of comoving isotropic (or "conformally flat") coordinates (Bolejko, Célérier & Krasinski 2011). Isotropic coordinates have the properties that the coordinate length of a small rigid rod is independent of the orientation of the rod and that the speed of light is independent of direction. So, isotropic coordinates are physically determined by local measurements using rigid scales or light triangulation (the radar method). Then the expansion parameter is used to convert to comoving coordinates.

The Stephani metric can be written

$$ds^2 = a^2(t)[k^2 dt^2 - \kappa^2(dx^2 + dy^2 + dz^2)] , \quad (12.1)$$

where $k$ and $\kappa$ depend on both time and position. Spherical symmetry is not assumed. $a$ is the expansion parameter, and can be assumed to satisfy Friedmann's equation in approximation. $k$ and $\kappa$ describe the gravitational perturbation. $k$ describes the gravitational redshift of light from gravitating body and gravitational potential in the weak field limit (or PPN formalism). $k$ describes potential troughs at the position of massive walls, near stationary in comoving coordinates and slowly increasing in depth as matter is accreted into the walls.

The calculation of $k$ and $\kappa$ depends on using either numerical solutions for Einstein's equation, or the solution in a neighbourhood (such as Schwarzschild geometry in isometric coordinates). However, since $k$ depends on an arbitrary (and unknown) mass distribution and variations in $\kappa$ can be ignored in the weak field limit, (12.1) can be simplified. We may usefully treat a single face of a buckyball universe in the PPN formalism using

$$ds^2 \approx a^2(t)[k^2 dt^2 - (dx^2 + dy^2 + dz^2)] , \quad (12.2)$$

In this approximation, $k$ can be determined from the Newtonian potential (noting that $k$ will depend upon $a$ in comoving coordinates). The result is that, in addition to recession velocities due to expansion, galaxies accelerate away from the centre of a void and towards the walls.

## 13 Lemaître's constant

In void models it is necessary to distinguish between the local value of Hubble's constant, $H_0 \approx 70 \text{ kms}^{-1}\text{Mpc}^{-1}$, describing the observed rate of recession, and the underlying, or background value, describing the rate of expansion. For clarity I will call the underlying value Lemaître's constant, since Lemaître identified the expansion of the universe while Hubble observed local expansion. Thus Lemaître's parameter is

$$L(t) = \frac{\dot{a}(t)}{a(t)} . \quad (13.1)$$

Lemaître's constant, $L_0 = L(t_0)$, is the current value of Lemaître's parameter) also called the background Hubble constant). A closed no-Λ cosmology, with a little over critical mass, assumed to lie in great walls beyond the horizon, and an age of 14 Gyrs has $L_0 \approx 45$ kms$^{-1}$Mpc$^{-1}$. This agrees with estimates for no-Λ cosmology based on the CMB by Biswas, Notari and Valkenburg (2010), and by Moss, Zibin and Scott (2011), but it should be recognised that those models were certainly oversimplified and did not describe the majority of mass residing in great walls extending to ~10 Myrs from the Big Bang.

Because a buckyball model places mass outside of the horizon, in walls where its gravitational effect is largely not detectable from within a cell, determination of Ω is almost certainly impossible. A significant difference between $H_0$ and $L_0$ is predicted, as matter tends to accelerate away from the centre of the void. As the difference depends on an unknown mass density profile as well as unknown initial conditions, one should not attach to it any fundamental importance. Based on a difference of 25 kms$^{-1}$Mpc$^{-1}$ between $H_0$ and $L_0$, one finds, by subtracting a Virgocentric velocity of 270 kms$^{-1}$ from the velocity of 627 kms$^{-1}$ for the local group relative to the CMB, that we are about 15 Mpc from the centre of the void, with some considerable uncertainty due to large expected peculiar velocities.

Moss, Zibin and Scott (2011) suggested that this violates a Copernican principle, that there are no special places in the universe. However, the Copernican principle allows variations due to the matter distribution. It is not violated by the fact that life is possible at the distance of the Earth from the Sun, but not on the solar surface or on Neptune. Similarly, the matter distribution generating a buckyball model is essentially arbitrary. Walls and voids form randomly, not at preferred positions. Life is not possible in superclusters and great walls because the background temperatures of the warm-hot intergalactic medium and of the intracluster medium are too great. Any observers must necessarily observe from within a void, but they may, in principle, observe from any part of the void. Observation tells us that we are at a particular location near the centre. This is not a *preferred* location. The probability of being at any particular point in space is zero; it does not follow that we cannot be at any point in space. This is simply an expression of the well-known property of the probability density function for a continuous distribution. There are many galaxies away from the centre of the void where life may equally be possible. Observers on those galaxies would measure a different value of the CMB dipole, but this has no particular significance. Each place (and motion) within a void allows only one value of the CMB dipole, but no particular value of the CMB dipole is preferred. It is neither coincidence nor violation of the cosmological principle to say that we are where we happen to be.

## 14 Summary of evidence

Standard cosmology offers no explanation for the observations of large voids and great walls or for alignments seen on all distance scales, from planar alignments of satellites, the Local Sheet, the galactic foam, large quasar and gamma ray burster groups. However, the formation of a cellular structure from baryonic collapse is well established and is described in textbooks, the only issue being the question of how baryonic collapse is initiated. After the start of collapse, the absence of a bound on the size of collapsing structure is a trivial consequence of general relativity (summarised crudely by saying that space itself collapses during gravitational collapse). Alignments on all distance scales are clearly predicted from the scale invariance of Jeans collapse. Baryonic collapse accounts for the sizes of observed structures and correctly predicts the approximately spherical form of observed large voids.

In addition to alignments in space, it is well established that members of galaxy groups, such as the Local Group, share peculiar motions which may be significantly different from those of neighbouring groups, denying a gravitational cause. Standard cosmology has offered no explanation. These common motions can only arise if the groups are formed from processes on greater

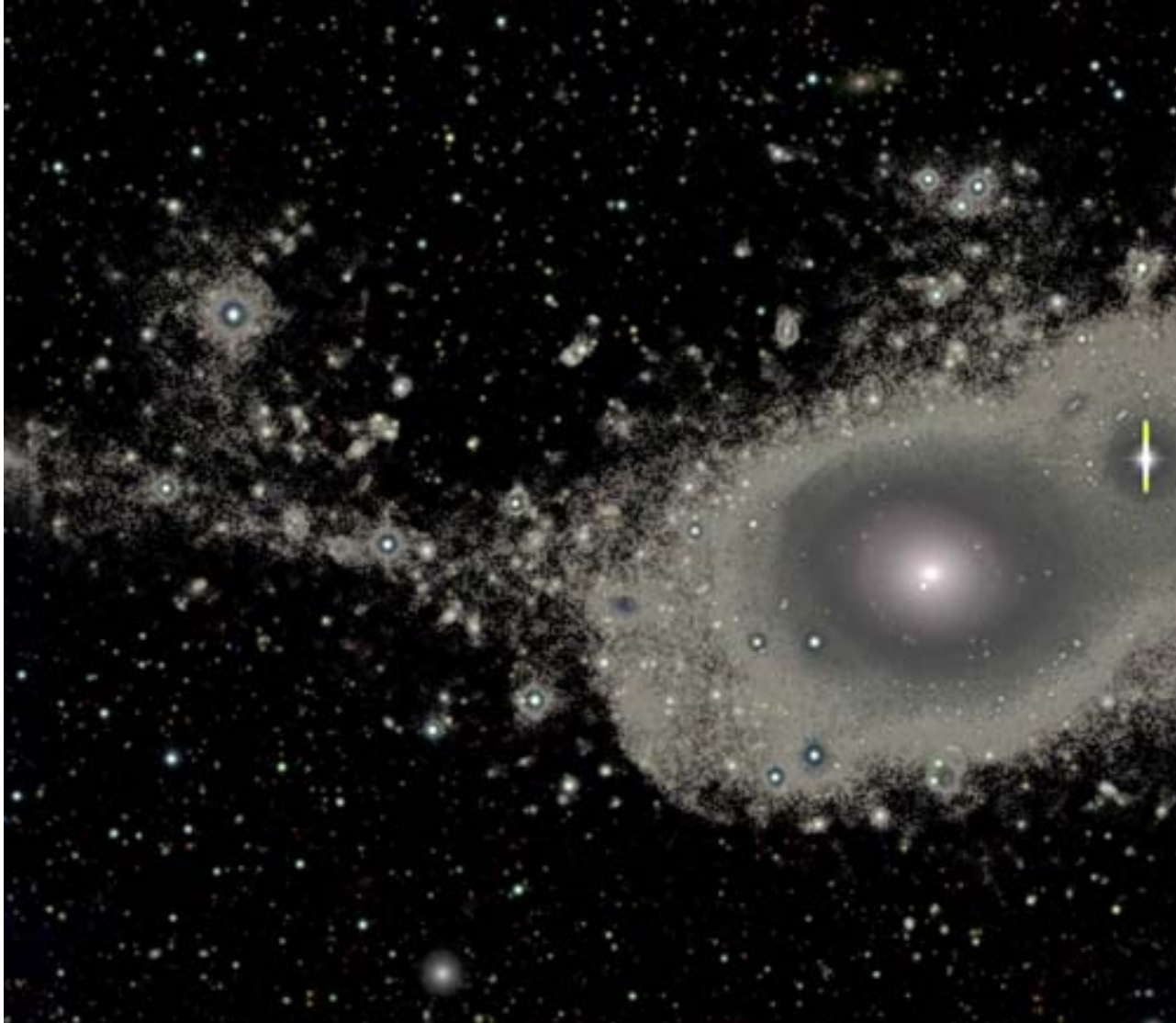

**Figure 22:** NGC 680 & NGC 5557 are members of a galaxy group about 40 million light years away. This image, by Pierre-Alain Duc & Jean-Charles Cuillandre using the Canada France Hawaii telescope, has been processed to enhance the shells surrounding the galaxies in this group.

distance scales, such as collisions between gaseous flows arising in the early universe.

Direct evidence of gas flows which can have only emanated from turbulence in the early universe is seen in regions of star formation where flows collide, for example in images of shell galaxies (e.g. NGC 680 & NGC 5557 figure 22, and NGC 474 figure 23). Shells are to be expected from star formation in high density regions where gas flowing inwards meets gas already present. Where gas streams collide, gas densities can become sufficient to trigger Jeans collapse and star formation. Wispy tails to galaxies are sometimes attributed to tidal forces, but tidal forces describe small differences in the accelerations of near bodies in free fall. As perturbations to gravity, tidal forces are extremely weak and slow acting, and cannot instigate gravitational collapse or star formation (similarly, the apparent force of the tides on Earth is the force of collisions following motions created over time).

Shells and tails can result from collisions between streams of gas, not unlike weather fronts. Similarly, the star-forming bridge of gas connecting the Small and the Large Magellanic Clouds, and the common envelope of neutral hydrogen surrounding them, shows collisions between gas flows generating a region of high pressure and density (Mathewson and Ford 1984, Heydari-Malayeri et al., 2003). The high space velocities of the Magellanic Clouds suggest that they are part of a system moving relative to the Milky Way (Kallivayalil et al. 2013), and also offers evidence of the residual motions of energetic gases emanating from the Big Bang.

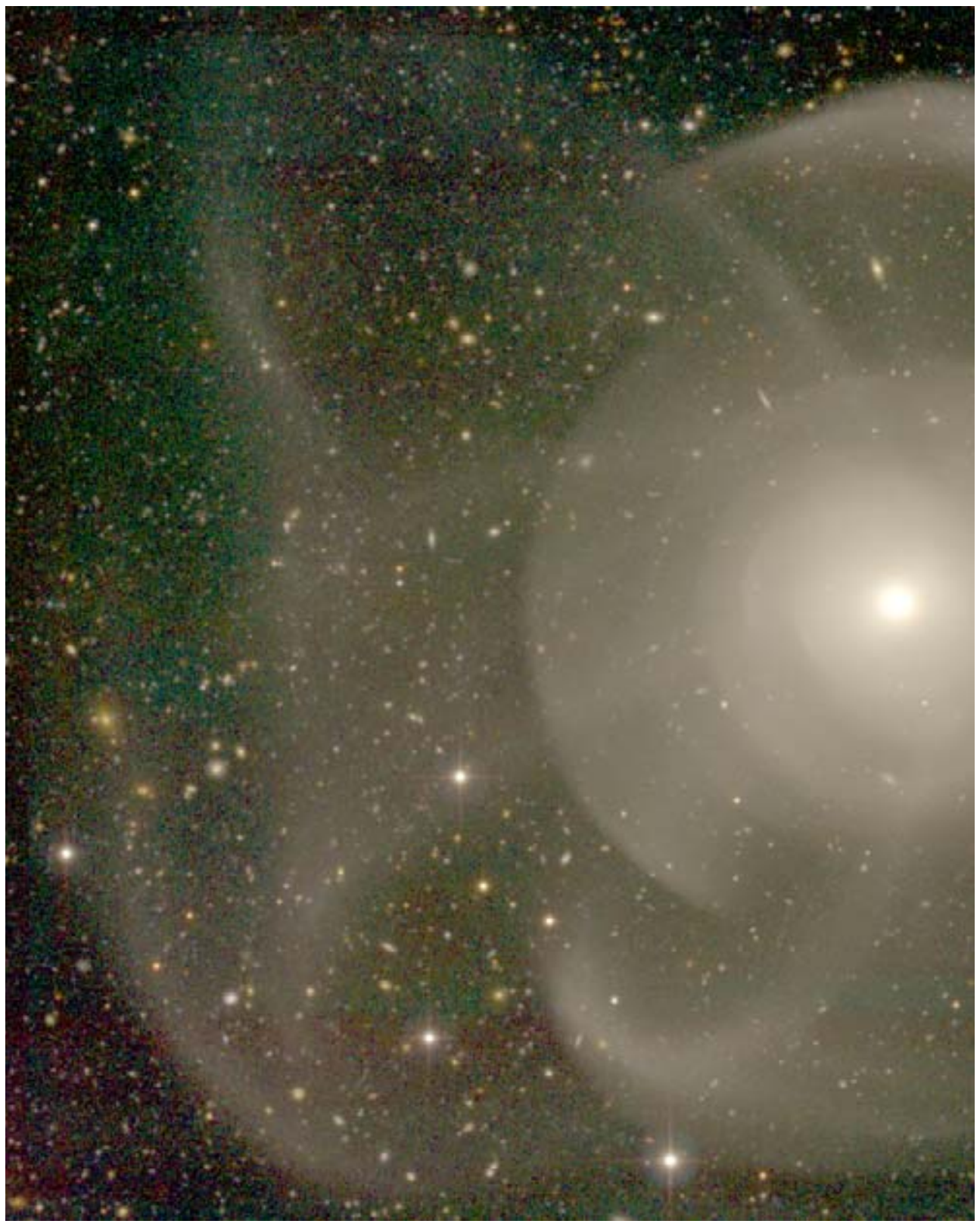

**Figure 23:** NGC 474 is about 80 kpc diameter, and about 100 million light years away. The shell structure becomes visible with high-quality imaging. Image by Pierre-Alain Duc & Jean-Charles Cuillandre with the Canada France Hawaii telescope.

Reionisation is not well explained in standard cosmology, but it is a clear consequence of the formation of the cellular structure predicted by baryonic collapse under gravity. The well established timescale for Jeans collapse shows that reionisation took place at precisely the time when it is expected that this structure should have formed.

The so-called "axis of evil" contradicts the prediction that the CMB is exactly homogeneous. The model described here expects small fluctuations in the CMB. It is trivially calculated that viscosity is so low and distances so large that a Reynolds number is many orders of magnitude greater than that needed to characterise a turbulent flow, and that turbulence will generate granularity on a scale much greater than that required to trigger Jeans collapse.

The original calculations of the abundances of light elements from BBN were made possible by using the simplifying assumption of uniform burning. This simplification is reasonable in approximation because mean behaviours in chaotic systems generally give similar results to behaviours in uniform systems (this is the usual method for treating turbulent systems). In practice, the lithium deficit shows that a correction is required to the standard calculation.

The first supposed evidence for dark matter was found in the 1930s by Fritz Zwicky, who used the virial theorem to observe that the Coma cluster could not be gravitationally bound in the absence of unseen matter. Given its size, the timescale for dispersion of the Coma cluster at 1000 $kms^{-1}$ is actually greater than the life of the universe. There is consequently no good reason to assume conditions of dynamic equilibrium in which the virial theorem can be applied. The motions of galaxies in the Coma cluster (and others) are more naturally taken to be direct evidence of high energies and turbulence in initial conditions.

The high incidence of abnormalities seen in galactic rotation curves directly refutes current explanations in terms of CDM and MOND, but is in clear agreement with evolution in a turbulent universe. The tendency to flatness agrees exactly with vortices formed in turbulent flows. The observed correlation between rotation speed and openness of spiral arms is as predicted, as is the location of spiral galaxies at the edges of cluster, and low surface brightness galaxies in sparse regions. The apparent greater dark matter content of low mass galaxies is accounted by the greater relative effect of surrounding gas. Measurements of gas at a distance of 1 Mpc confirm the presence of gas streams independent of the gravity of the Milky Way.

Galactic rotation curves have been thought to offer evidence of CDM in spiral galaxies, but this paper has shown that the Navier-Stokes equation leads to the formation of spirals due to collisions in intergalactic gas, and that N-body simulations confirm the role of spiral potential on stellar velocities. The lack of apparent dark matter in elliptical galaxies, not explained in standard cosmology, is accounted because ellipticals are not formed from vortices in turbulent flows.

## 15 Conclusion

It has been widely suggested that cosmological observations provide evidence for exotic physics, not observed on Earth and for which there is no proper rationale in terms of known physical law. However, this paper has shown that arguments for the stability of a homogeneous initial condition violate the continuity equation for mass energy and the principle that local dynamics depends on local energy differences, not on the high energy of the background.

Using straightforward arguments in conventional physics, this paper has observed from the well understood processes of fusion and the behaviours of gases and radiation that the homogeneity of the early universe is necessarily unstable, generating turbulence and leading naturally to observed structure. Based on the known physics of baryonic structure formation and using the calculated properties of the Jeans mechanism and Newton's shell theorem, it is possible to account for a wide range of observational data for which there has been little or no explanation in standard cosmology. This data constitutes significant quantitative evidence for turbulence in the early universe, and includes structures observed on all distance scales, peculiar motions of galaxy groups, the general form of, and abnormalities in, galaxy rotation curves, the lack of apparent dark matter in elliptical galaxies, the apparent high dark matter content of low mass spiral galaxies, high velocity dispersions in galaxy clusters, reionisation, the flatness problem and the quantitative fit of cosmological parameters with the CMB in void models.

General relativity shows that there is no bound on the size of structures formed from gravitational collapse. It has been seen that in a baryonic universe with a homogeneous initial condition, Big Bang nucleosynthesis would rapidly lead to a universe with a fractal Voronoi, or "buckyball" space geometry, generating observed structures on all distance scales. A buckyball geometry offers a natural resolution of the "flatness problem" without recourse to non-standard physics, since mass is concentrated in walls beyond the horizon.

In void models, the cosmological principle is obeyed in approximation on large enough distance scales. Geometry can be described by perturbing Friedmann cosmology using Stephani metric, in accordance with the expected consequence of Jeans collapse, and with important implications for the calculation of cosmological parameters from observational data. It is to be expected that the current value of Hubble's constant substantially overestimates the true rate of expansion. Published calculations for void models have shown that have shown that it is not possible to reject a no-Λ, no-CDM cosmology on grounds of the fit to the spectrum of the CMB. The cosmic microwave background is consistent with an underlying Hubble constant, or Lemaître constant, of $L_0 \approx$ 45 $\text{kms}^{-1}\text{Mpc}^{-1}$. This rate of expansion is consistent with a no-Λ universe with a little over critical mass (assumed to be in great walls beyond the horizon) and an age of ~14 Gyrs. The calculation of a cosmological constant is seen as an artefact of the acceleration of distant galaxies away from the centre of a void and towards regions of greater density in all directions.

While standard cosmology invokes non-standard physics, the question whether observation permits interpretation in terms of standard physics remains of great importance. We must conclude that it is not possible at the current time to reject explanations in conventional physics for observational data.